# Robust Bayesian compressive sensing for signals in structural health monitoring

Yong Huang

*Division of Engineering and Applied Science, California Institute of Technology, Pasadena, CA 91125, USA
and School of Civil Engineering, Harbin Institute of Technology, Harbin 150090, China*

James L. Beck* & Stephen Wu

*Division of Engineering and Applied Science, California Institute of Technology, Pasadena, CA 91125, USA*

&

Hui Li

*School of Civil Engineering, Harbin Institute of Technology, Harbin 150090, China*

**Abstract:** *In structural health monitoring (SHM) systems for civil structures, massive amounts of data are often generated that need data compression techniques to reduce the cost of signal transfer and storage, meanwhile offering a simple sensing system. Compressive sensing (CS) is a novel data acquisition method whereby the compression is done in a sensor simultaneously with the sampling. If the original sensed signal is sufficiently sparse in terms of some orthogonal basis (e.g. a sufficient number of wavelet coefficients are zero or negligibly small), the decompression can be done essentially perfectly up to some critical compression ratio; otherwise there is a trade-off between the reconstruction error and how much compression occurs. In this article, a Bayesian compressive sensing (BCS) method is investigated that uses sparse Bayesian learning to reconstruct signals from a compressive sensor. By explicitly quantifying the uncertainty in the reconstructed signal from compressed data, the BCS technique exhibits an obvious benefit over existing regularized norm-minimization CS methods that provide a single signal estimate. However, current BCS algorithms suffer from a robustness problem: sometimes the reconstruction errors are very large when the number of measurements K is a lot less than the number of signal degrees of freedom N that are needed to capture the signal accurately in a directly sampled form. In this paper, we present improvements to the BCS reconstruction method to enhance its robustness so that even higher compression ratios N/K can be used and we examine the tradeoff between efficiently compressing data and accurately decompressing it. Synthetic data and actual acceleration data collected from a bridge SHM system are used as examples. Compared with the state-of-the-art BCS reconstruction algorithms, the improved BCS algorithm demonstrates superior performance. With the same acceptable-error rate based on a specified threshold of reconstruction error, the proposed BCS algorithm works with relatively large compression ratios and it can achieve perfect lossless compression performance with quite high compression ratios. Furthermore, the error bars for the signal reconstruction are also quantified effectively.*

## 1 INTRODUCTION

Structural health monitoring (SHM) for civil engineering infrastructure refers to the process of implementing a damage detection and characterization strategy based on data from sensors distributed over a system (Beck et al., 2001; Gangone et al., 2011; Jiang and Adeli, 2007; Lynch, 2007; Sohn et al., 2003; Vanik et al., 2000). A substantial

---

*To whom correspondence should be addressed. E-mail: *jimbeck@caltech.edu*.

number of sensors are required for SHM systems due to the complexity and large scale of civil structures. Consequently, a large amount of data is usually produced by these systems, especially those that are continuously monitoring large civil structures such as long span suspension and cable-stayed bridges. Therefore, data compression is needed to reduce the cost and increase the efficiency of signal transfer and storage.

Data compression for SHM systems has attracted much interest in recent years, especially for wireless monitoring systems, since data compression techniques can provide a way to improve the power efficiency and minimize bandwidth during the transmission of structural response time-histories from wireless sensors (Lynch et al., 2003; Lynch, 2007; Xu et al., 2004). Wavelet-based compression techniques (Xu et al., 2004) and Huffman lossless compression techniques (Lynch et al., 2003) have been developed. All of these data compression methods belong to a conventional framework for sampling signals that follow the Nyquist-Shannon theorem: the sampling rate must be at least twice the maximum frequency present in the signal.

Compressive sensing (CS) (Candes and Wakin, 2008; Donoho, 2006) is a novel sampling technique for data acquisition, whose capability, when first met, seems surprising. It asserts that if certain signals are sparse in some orthogonal basis, one can reconstruct these signals accurately from far fewer measurements than what is usually considered necessary based on Nyquist-Shannon sampling. This new technique may become the main paradigm for sampling and compressing data simultaneously, therefore increasing the efficiency of data transfer and storage. The basic idea for data sampling and compressing in CS is to use a specially-designed sensor (e.g. Yoo et al., (2012)) to project $N$ time-sampled points of a sensed signal to $K \ll N$ compressed data points using a randomly chosen $K \times N$ projection matrix. The second stage for CS refers to the data reconstruction or decompression, which is performed after the compressed data is transmitted to the data management center. A least-squares method with $l_1$–norm regularization (often called Basis Pursuit (Candes et al., 2006; Chen et al., 1999)) is then used to reconstruct the original signal accurately based on the $K$ compressed measurements. This reconstruction problem has attracted the interest of many researchers and, in recent years, alternatives to the original CS reconstruction method have been proposed, such as Orthogonal Matching Pursuit (Tropp and Gilbert, 2007) and Bayesian Compressive Sensing (BCS) (Ji et al., 2008).

In this article, as in Ji et al. (2008), we apply sparse Bayesian learning for the CS reconstruction problem but we focus on the robustness of this procedure. In contrast to the original CS reconstruction algorithms (Candes et al., 2006; Chen et al., 1999; Tropp and Gilbert, 2007) that provide only a point estimate of the basis coefficients to specify the signal reconstruction, the BCS algorithm uses posterior probability distributions over these coefficients as an efficient way to quantify uncertainty in the reconstructed signals. However, we show that current BCS reconstruction algorithms suffer from a robustness problem: sometimes when the compression ratio is high, these iterative algorithms converge to suboptimal solutions with very large reconstruction errors. Therefore it is desirable to develop a robust reconstruction algorithm that allows higher compression ratios.

An improved BCS algorithm is proposed here to reduce the likelihood of sub-optimal solutions of the optimization problem during the reconstruction process and so to produce more robustly smaller reconstruction errors. We demonstrate with experimental results from synthetic data and actual data from a SHM system on a bridge that the proposed algorithm provides better reconstruction performance than the state-of-the-art BCS methods, both for signal reconstruction robustness and uncertainty quantification.

## 2 BAYESIAN COMPRESSIVE SENSING

Consider a discrete-time signal $\mathbf{x} = [x(1), \cdots x(N)]^T$ in $\mathbb{R}^N$ represented by a set of orthogonal basis vectors as

$$\mathbf{x} = \sum_{n=1}^{N} w_n \mathbf{\Psi}_n \text{ or } \mathbf{x} = \mathbf{\Psi}\mathbf{w} \qquad (1)$$

where $\mathbf{\Psi} = [\mathbf{\Psi}_1, \cdots, \mathbf{\Psi}_N]$ is the $N \times N$ basis matrix with the orthonormal basis of $N \times 1$ vectors $\{\mathbf{\Psi}_n\}_{n=1}^{N}$ as columns; $\mathbf{w}$ is the sparse coefficients or weight vector, i.e., it is known that most of its components are zero or very small (with minimal impact on the signal) but not which ones. The total number of the zero components of $\mathbf{w}$ represents the *sparsity* of the signal $\mathbf{x}$ with respect to the basis $\{\mathbf{\Psi}_n\}$ (Wipf and Rao, 2006).

In the framework of CS, one infers the coefficients $w_n$ of interest from compressed data instead of directly sampling the actual signal $\mathbf{x}$. The data vector $\mathbf{y}$ from the compressive sensor is composed of $K \ll N$ linear projections of the signal $\mathbf{x}$ using a chosen $K \times N$ random projection matrix $\mathbf{\Phi}$ that is built into the sensor (each element in $\mathbf{\Phi}$ is an independent sample drawn from $\mathcal{N}(0,1)$):

$$\mathbf{y} = \mathbf{\Phi}\mathbf{x} + \mathbf{r} \qquad (2)$$

where $\mathbf{r}$ represents any measurement error. For reconstruction, (1) and (2) are combined to represent the compressed data $\mathbf{y}$ as:

$$\mathbf{y} = \mathbf{\Theta}\mathbf{w} + \mathbf{e} = \sum_{n=1}^{N} w_n \mathbf{\Theta}_n + \mathbf{e} \qquad (3)$$

where $\mathbf{\Theta} = \mathbf{\Phi}\mathbf{\Psi}$ and $\mathbf{e}$ represents the unknown prediction error due to the signal model for specified $\mathbf{w}$ plus the measurement error $\mathbf{r}$. Because $\mathbf{\Theta}$ is a $K \times N$ matrix with $K \ll N$, (3) leads to an ill-posed inversion problem for finding the weights $\mathbf{w}$, and hence the signal $\mathbf{x}$ in $\mathbb{R}^N$, from data $\mathbf{y}$ in $\mathbb{R}^K$.

By exploiting the sparsity of the representation of **x** in basis $\{\Psi_n\}_{n=1}^N$, the ill-posed problem can be posed as a convex least-squares regularization problem to estimate **w** as follows (Chen et al., 1999; Candes et al., 2006; Tropp and Gilbert, 2007):

$$\widetilde{\mathbf{w}} = \arg\ \min\{\|\mathbf{y} - \mathbf{\Theta w}\|_2^2 + \lambda\|\mathbf{w}\|_1\} \quad (4)$$

where the penalty parameter $\lambda$ scales the regularization term to penalize large weight values. As a result, the optimization problem in (4) represents a trade-off between how well the data is fitted (first term) and how sparse the signal is (second term). Appropriate minimization algorithms have been proposed, including linear programming in Basis Pursuit (Chen et al., 1999; Candes et al., 2006) and greedy algorithms in Orthogonal Matching Pursuit (Tropp and Gilbert, 2007), to solve the CS reconstruction problem as formulated in (4). The choice of a 1-norm for the regularization in (4) is important because it induces sparsity in $\widetilde{\mathbf{w}}$, while still giving a convex optimization problem.

### 2.1 Sparse Bayesian learning of linear regression model for compressive sensing

The ill-posed data inversion problem can also be tackled using a Bayesian perspective, which has certain distinct advantages; for example, in addition to providing a sparse solution to estimate the underlying signal, it automatically estimates penalty parameters like $\lambda$ in (4) and it provides a measure of the uncertainty for the reconstructed signal. The basic idea is to use sparse Bayesian learning (Tipping, 2001; Tipping and Faul, 2003) for the linear regression problem involved in CS reconstruction (Ji et al., 2008). Therefore, Bayes' theorem is applied to find the posterior probability density function (PDF) $p(\mathbf{w}|\mathbf{y})$ for the signal weights **w** in (3) based on the linearly projected data **y**. The error **e** in (3) is modeled as a zero-mean Gaussian vector with covariance matrix $\sigma^2 \mathbf{I}_K$. This probability model gives the largest uncertainty for **e** (i.e. maximizes Shannon's information entropy (Jaynes, 1957)) subject to the first two moment constraints: $\mathbf{E}[e_k] = 0, \mathbf{E}[e_k^2] = \sigma^2, k = 1, \dots, K$. Thus, one gets a Gaussian likelihood function:

$$p(\mathbf{y}|\mathbf{w}, \sigma^2) = (2\pi\sigma^2)^{-\frac{K}{2}} \exp\left(-\frac{1}{2\sigma^2}\|\mathbf{y} - \mathbf{\Theta w}\|_2^2\right) \quad (5)$$

This likelihood measures how well the signal model for specified parameters **w** and $\sigma^2$ predicts the observed CS measurements **y**, and it corresponds to the first term of (4) in the deterministic CS data inversion.

For the Bayesian prior distribution on the model parameters, independent prior PDFs are taken for $\sigma^2$ and **w**. A Gamma prior is placed on $\sigma^{-2}$:

$$p(\sigma^{-2}) = \Gamma(\sigma^{-2}|a, b) = \frac{b^a}{\Gamma(a)}(\sigma^{-2})^{a-1}\exp(-b\sigma^{-2}) \quad (6)$$

Based on the sparse Bayesian learning approach, a special Gaussian prior distribution is used for the signal coefficients **w**, which is known as the *automatic relevance determination prior* (ARD prior):

$$\begin{aligned} p(\mathbf{w}|\boldsymbol{\alpha}) &= \prod_{n=1}^N p(w_n|\alpha_n) \\ &= \prod_{n=1}^N \left[(2\pi)^{-1/2}\alpha_n^{1/2}\exp\left\{-\frac{1}{2}\alpha_n w_n^2\right\}\right] \end{aligned} \quad (7)$$

where the hyperparameter $\alpha_n$ is the precision (inverse prior variance) for $w_n$. It is used to control the model sparsity by having an effect similar to the regularization term in (4). Actually, the equivalent Bayesian formulation to the optimization in (4) is to find the MAP (maximum a posteriori) value of **w** using the likelihood function in (5) with a Laplace prior $p(\mathbf{w}|\alpha_0) = \alpha_0\exp(-\alpha_0\|\mathbf{w}\|_1)$, where $\alpha_0 = \lambda/2\sigma^2$. For this choice, however, the full Bayesian posterior distribution for **w** and **x** cannot be found analytically, in contrast to the ARD prior, which corresponds to using a weighted 2-norm on **w** in (4). Using the 2-norm in (4), however, turns out to not promote sparsity when estimating **w**. To induce sparsity, in sparse Bayesian learning (Tipping, 2001; Tipping and Faul, 2003) we optimize over the hyperparameter vector $\boldsymbol{\alpha}$ to find its MAP value (most probable value based on the data **y**).

Given the CS measurements **y** and the prior on **w**, the posterior distribution $p(\mathbf{w}|\mathbf{y}, \boldsymbol{\alpha}, \sigma^2)$ over the weights is obtained based on Bayes' theorem:

$$p(\mathbf{w}|\mathbf{y}, \boldsymbol{\alpha}, \sigma^2) = p(\mathbf{y}|\mathbf{w}, \sigma^2)p(\mathbf{w}|\boldsymbol{\alpha})/p(\mathbf{y}|\boldsymbol{\alpha}, \sigma^2) \quad (8)$$

where $p(\mathbf{w}|\boldsymbol{\alpha})$ = prior PDF of **w** in (7); $p(\mathbf{y}|\mathbf{w}, \sigma^2)$ = likelihood function in (5); and $p(\mathbf{y}|\boldsymbol{\alpha}, \sigma^2)$ is the *evidence* of the signal model class $\mathcal{M}(\boldsymbol{\alpha}, \sigma^2)$ which also serves as a normalizing constant for the posterior distribution. Since both prior and likelihood for **w** are Gaussian and the likelihood mean $\mathbf{\Theta w}$ is linear in **w**, the posterior PDF can be expressed analytically as a multivariate Gaussian distribution $p(\mathbf{w}|\mathbf{y}, \boldsymbol{\alpha}, \sigma^2) = N(\boldsymbol{\mu}, \boldsymbol{\Sigma})$ with mean and $N \times N$ covariance matrix (Tipping, 2001; Tipping and Faul, 2003):

$$\boldsymbol{\mu} = \sigma^{-2}\boldsymbol{\Sigma}\mathbf{\Theta}^T\mathbf{y} \quad (9)$$

$$\boldsymbol{\Sigma} = (\sigma^{-2}\mathbf{\Theta}^T\mathbf{\Theta} + \mathbf{A})^{-1} \quad (10)$$

where $\mathbf{A} = \text{diag}(\alpha_1, \dots, \alpha_N)$. It follows from (1) that for given hyperparameters $\boldsymbol{\alpha}$ and $\sigma^2$ defining the model class $\mathcal{M}(\boldsymbol{\alpha}, \sigma^2)$, the posterior probability distribution for the reconstructed signal **x** is Gaussian, $p(\mathbf{x}|\mathbf{y}, \boldsymbol{\alpha}, \sigma^2) = N(\boldsymbol{\Psi\mu}, \boldsymbol{\Psi\Sigma\Psi}^T)$.

In the next step, Bayesian model class assessment (Beck, 2010) is used to select the MAP value of hyperparameters $\boldsymbol{\alpha}$ and $\sigma^2$. If the problem is globally identifiable (Beck and Katafygiotis, 1998), meaning here that the evidence $p(\mathbf{y}|\boldsymbol{\alpha}, \sigma^2)$ has a single pronounced global maximum with respect to $\boldsymbol{\alpha}$ and $\sigma^2$, then reconstruction can be done accurately using the most probable model class $\mathcal{M}(\widehat{\boldsymbol{\alpha}}, \widehat{\sigma}^2)$ based on measurements $\mathbf{y}$ where $\hat{\boldsymbol{\alpha}}$ and $\hat{\sigma}^2$ are the MAP values that maximize $p(\boldsymbol{\alpha}, \sigma^2 | \mathbf{y}) \propto p(\mathbf{y}|\boldsymbol{\alpha}, \sigma^2) p(\boldsymbol{\alpha}) p(\sigma^2)$. The mathematical justification for using the MAP values rather than the full predictive PDF is that it can be approximated by applying Laplace's asymptotic approximation (Beck and Katafygiotis, 1998; Beck, 2010):

$$p(\mathbf{x}|\mathbf{y}) = \int p(\mathbf{x}|\mathbf{y}, \boldsymbol{\alpha}, \sigma^2) p(\boldsymbol{\alpha}, \sigma^2|\mathbf{y}) d\boldsymbol{\alpha} d\sigma^2 \approx p(\mathbf{x}|\mathbf{y}, \hat{\boldsymbol{\alpha}}, \hat{\sigma}^2) \quad (11)$$

By taking a uniform prior on $\boldsymbol{\alpha}$ and inverse Gamma prior on $\sigma^2$, maximization of the following objective function is required:

$$\begin{aligned} p(\boldsymbol{\alpha}, \sigma^2|\mathbf{y}) &= p(\mathbf{y}|\boldsymbol{\alpha}, \sigma^2) p(\boldsymbol{\alpha}) p(\sigma^{-2}) \\ &= \int p(\mathbf{y}|\mathbf{w}, \sigma^2) p(\mathbf{w}|\boldsymbol{\alpha}) p(\boldsymbol{\alpha}) p(\sigma^{-2}) d\mathbf{w} \\ &= \frac{b^a}{\Gamma(a)} (2\pi)^{-\frac{K}{2}} (\sigma^{-2})^{a-1} |\sigma^2 \mathbf{I} + \boldsymbol{\Theta} \mathbf{A}^{-1} \boldsymbol{\Theta}^T|^{-\frac{1}{2}} \\ &\quad \exp\left\{-b\sigma^{-2} - \frac{1}{2}\mathbf{y}^T(\sigma^2\mathbf{I} + \boldsymbol{\Theta}\mathbf{A}^{-1}\boldsymbol{\Theta}^T)^{-1}\mathbf{y}\right\} \end{aligned} \quad (12)$$

Since $p(\boldsymbol{\alpha})$ is constant over the important region of the $\boldsymbol{\alpha}$ space, the optimization over $\boldsymbol{\alpha}$ for given $\sigma^2$ is equivalent to maximizing the evidence function $p(\mathbf{y}|\boldsymbol{\alpha}, \sigma^2) = \int p(\mathbf{y}|\mathbf{w}, \sigma^2) p(\mathbf{w}|\boldsymbol{\alpha}) d\mathbf{w}$. We now briefly describe two optimization algorithms that have been proposed to find the MAP values $\hat{\boldsymbol{\alpha}}$ and $\hat{\sigma}^2$. One is Tipping's original iterative algorithm (Tipping, 2001), which we call the "Top-down" algorithm. Another one is Tipping and Faul's "Fast Algorithm" (Tipping and Faul, 2003), which we call the "Bottom-up" algorithm.

### 2.2 Algorithms for estimation of the optimal hyperparameters $\boldsymbol{\alpha}$ and $\sigma^2$

*Top-down algorithm.* The Top-down algorithm starts with considering all basis vectors in expansion (1) and (3) and tries to prune out irrelevant basis vectors by maximizing the evidence $p(\mathbf{y}|\boldsymbol{\alpha}, \sigma^2)$ which Tipping has shown automatically yields sparsity in the expansion (Tipping, 2001; Tipping and Faul, 2003). By setting derivatives of the log evidence $\ln p(\mathbf{y}|\boldsymbol{\alpha}, \sigma^2)$ with respect to $\boldsymbol{\alpha}$ to zero, the following iterative estimation formula is obtained for each $n$ (see Tipping (2001)):

$$\alpha_n^{new} = \frac{1 - \alpha_n \Sigma_{nn}}{\mu_n^2} \quad (13)$$

where $\mu_n$ and $\Sigma_{nn}$ are computed from (9) and (10) using the current $\boldsymbol{\alpha}$ and $\sigma^2$. For $\sigma^2$, setting the derivative of the logarithm of (12) with respect to $\sigma^2$ equal to zero and rearranging, gives (see Appendix A):

$$\sigma^2 = \frac{\|\mathbf{y} - \boldsymbol{\Theta}\boldsymbol{\mu}\|^2 + 2b}{K - \sum_{n=1}^{N'} (1 - \alpha_n \Sigma_{nn}) + 2(a-1)} \quad (14)$$

where $N'$ is the number of non-zero terms in (1). Additionally, we also find the optimal values of the parameters $b$ and $a$ in (6) by taking the derivative of the logarithm of (12) with respect to $b$ and $a$, giving (see Appendix A):

$$b = a\sigma^2 \quad (15)$$

$$\log b - \psi(a) - \log \sigma^2 = 0 \quad (16)$$

where $\psi(a)$ is the Digamma function. In principle, the process iterates between the hyperparameters ($\boldsymbol{\alpha}$ and $\sigma^2, b, a$) and the posterior parameters ($\boldsymbol{\mu}$ and $\boldsymbol{\Sigma}$), until a convergence criterion has been satisfied to obtain the optimal hyperparameters.

In this algorithm, the $n^{th}$ basis function is deleted when the optimal $\alpha_n$ approaches infinity during the iteration process. In practice, many of the $\alpha_n$ approach infinity, which implies that their corresponding coefficients have insignificant amplitudes for representation of the measurements $\mathbf{y}$ because each such $w_n$ has a zero-mean Gaussian prior with variance approaching zero. Since the number of retained coefficients is usually quite small compared with the length of data, a sparsely reconstructed signal model is obtained. One drawback of the method is that basis vectors are pruned when individual iterated $\alpha_n$ values grow numerically too large, which is not a precise analytical procedure to produce sparsity of signal models. Also, this algorithm involves inversion of matrices of size $N \times N$, thereby making this approach relatively slow for reconstruction of signals of large dimension; however, for sparse signal reconstruction from overcomplete dictionaries such as compressive sensing decompression ($K \ll N$), we can use the Woodbury inversion identity as follows (Tipping, 2001):

$$\boldsymbol{\Sigma} = (\sigma^{-2}\boldsymbol{\Theta}^T\boldsymbol{\Theta} + \mathbf{A})^{-1} = \mathbf{A}^{-1} - \mathbf{A}^{-1}\boldsymbol{\Theta}^T \mathbf{C}^{-1} \boldsymbol{\Theta} \mathbf{A}^{-1} \quad (17)$$

where $\mathbf{C} = \sigma^2\mathbf{I} + \boldsymbol{\Theta}\mathbf{A}^{-1}\boldsymbol{\Theta}^T$. We now reduce the algorithm to an inversion of matrix $\mathbf{C}$ with $O(K^3)$ complexity, which is clearly superior when $K < N'$.

*Bottom-up algorithm.* The Bottom-up algorithm starts with no terms in (1) and adds relevant ones to the signal model as the iterations proceed. This method can significantly reduce the reconstruction time and the chance of having ill-conditioning problems during inversion of the matrix in (10). In this algorithm, the log evidence $\log p(\mathbf{y}|\boldsymbol{\alpha}, \sigma^2)$ is expressed as

$$\begin{aligned} \mathcal{L}(\boldsymbol{\alpha}, \sigma^2) &= \log p(\mathbf{y}|\boldsymbol{\alpha}, \sigma^2) \\ &= -\frac{1}{2}[K\log 2\pi + \log|\mathbf{C}| + \mathbf{y}^T \mathbf{C}^{-1}\mathbf{y}] \\ &= -\frac{1}{2}\Big[K\log 2\pi + \log|\mathbf{C}_{-n}| + \mathbf{y}^T \mathbf{C}_{-n}^{-1}\mathbf{y} - \log \alpha_n \\ &\quad + \log(\alpha_n + \boldsymbol{\Theta}_n^T \mathbf{C}_{-n}^{-1} \boldsymbol{\Theta}_n) - \frac{(\boldsymbol{\Theta}_n^T \mathbf{C}_{-n}^{-1}\mathbf{y})^2}{\alpha_n + \boldsymbol{\Theta}_n^T \mathbf{C}_{-n}^{-1}\boldsymbol{\Theta}_n}\Big] \\ &= \mathcal{L}(\boldsymbol{\alpha}_{-n}, \sigma^2) + \frac{1}{2}\Big[\log \alpha_n - \log(\alpha_n + \boldsymbol{\Theta}_n^T \mathbf{C}_{-n}^{-1}\boldsymbol{\Theta}_n) \\ &\quad + \frac{(\boldsymbol{\Theta}_n^T \mathbf{C}_{-n}^{-1}\mathbf{y})^2}{\alpha_n + \boldsymbol{\Theta}_n^T \mathbf{C}_{-n}^{-1}\boldsymbol{\Theta}_n}\Big] \\ &= \mathcal{L}(\boldsymbol{\alpha}_{-n}, \sigma^2) + \frac{1}{2}\Big[\log \alpha_n - \log(\alpha_n + S_n) + \frac{Q_n^2}{\alpha_n + S_n}\Big] \\ &= \mathcal{L}(\boldsymbol{\alpha}_{-n}, \sigma^2) + l(\alpha_n, \sigma^2) \end{aligned} \quad (18)$$

where $\mathbf{C} = \sigma^2\mathbf{I} + \mathbf{\Theta}\mathbf{A}^{-1}\mathbf{\Theta}^T$ and $\mathbf{C}_{-n}$ = covariance matrix $\mathbf{C}$ with the components of $n$ removed, and therefore $\mathcal{L}(\boldsymbol{\alpha}_{-n}, \sigma^2)$ does not depend on $\alpha_n$. In addition, for simplification of forthcoming expressions, we have defined the 'sparsity factor' $S_n$ and 'quality factor' $Q_n$ by:

$$S_n = \mathbf{\Theta}_n^T \mathbf{C}_{-n}^{-1} \mathbf{\Theta}_n \tag{19}$$

$$Q_n = \mathbf{\Theta}_n^T \mathbf{C}_{-n}^{-1} \mathbf{y}. \tag{20}$$

Setting derivatives of (18) with respect to $\alpha_n$ to zero leads to

$$\alpha_n = \begin{cases} \infty, & \text{if } Q_n^2 \leq S_n \\ \frac{S_n^2}{Q_n^2 - S_n}, & \text{if } Q_n^2 > S_n \end{cases} \tag{21}$$

This algorithm enables an efficient sequential optimization by updating one candidate basis term at each iteration to monotonically increase the evidence. Finally, only the components that have finite $\alpha_n$ are used in determining the signal model since each $w_n$ with $\alpha_n = \infty$ has prior mean and variance both zero, giving $w_n = 0$, and so its term drops out of (1) and (3). For updating $\sigma^2$, the same equation as in (14) can be used for the Bottom-up algorithm.

When calculating $S_n$ and $Q_n$ in (19) and (20) for updating $\alpha_n$, the Bottom-up algorithm requires the numerical inversion of the $K \times K$ matrix $\mathbf{C}$, which is an $O(K^3)$ operation. The Woodbury inversion identity can be employed:

$$\mathbf{C}^{-1} = (\sigma^2\mathbf{I} + \mathbf{\Theta}\mathbf{A}^{-1}\mathbf{\Theta}^T)^{-1} = \sigma^{-2}\mathbf{I} - \sigma^{-2}\mathbf{\Theta}\mathbf{\Sigma}\mathbf{\Theta}^T\sigma^{-2} \tag{22}$$

where the $N' \times N'$ covariance matrix $\mathbf{\Sigma}$, which is given in (10), is readily computed in the optimization procedure; it depends on the size $N'$ of the current signal models, which is much smaller than $K$ for the Bottom-up optimization procedure. Therefore, the Bottom-up method has an obvious advantage over the Top-down algorithm in terms of algorithm efficiency.

**2.3 The robustness problem of BCS reconstruction**

*Effect of number of measurements on reconstruction.* In the fully probabilistic framework of BCS, the most plausible signal reconstruction of $\mathbf{x}$ is the posterior mean $\hat{\mathbf{x}} = \mathbf{\Psi}\boldsymbol{\mu}$ where the posterior mean $\boldsymbol{\mu}$ of $\mathbf{w}$ is given by (9) with the optimal hyperparameters $\hat{\boldsymbol{\alpha}}$ and $\hat{\sigma}^2$. For brevity, we call $\hat{\mathbf{x}}$ the reconstructed signal. The number of measurements $K$ in the compressed data $\mathbf{y}$ has an important influence on the robustness of signal reconstruction. However, in order to compress the signal more effectively, the number of measurements must be reduced to be much smaller than the number of degrees of freedom $N$ of the original signal $\mathbf{x}$. As a result, reconstruction of the signal becomes a severely ill-posed problem that sometimes leads to sub-optimal signal representations during the optimization over the hyperparameters. This occurs because there are a large number of local maxima that trap the optimization and significantly reduce the robustness of the iterative scheme, for both the Top-down and Bottom-up algorithms.

To illustrate the lack of robustness, we apply Bottom-up and Top-down algorithms to synthetic data considering of 1000 different random measurement samples for each specific size $K$ (by changing the random projection matrix $\mathbf{\Phi}$ each time) but using the same original signal $\mathbf{x}$ of length $N = 512$ each time. Two signals are considered, each containing 20 non-zero spikes created by randomly choosing 20 discrete times and randomly drawing the amplitudes of the spikes from two different probability distributions, one uniform with equal probability of $\pm 1$ random spikes and the other a zero-mean unit variance Gaussian. For reconstruction, we take $\mathbf{\Psi} = \mathbf{I}_N$ (identity matrix), so $\mathbf{\Theta} = \mathbf{\Phi}$. The data is noise free but for reconstruction, we fix the prediction error variance to be $\sigma^2 = \text{var}[\mathbf{y}] \times 0.1$. To investigate the effect of the number of measurements $K$ on reconstruction, we present the reconstruction error, log evidence and the size of the final reconstructed signal models (the number of non-zero terms in (1)) as a function of the number of measurements $K$ which has values from 10 to 520 in Figures 1 and 2, for Bottom-up and Top-down algorithms, respectively. The reconstruction error is defined as $\|\hat{\mathbf{x}} - \mathbf{x}\|_2^2 / \|\mathbf{x}\|_2^2$, where $\hat{\mathbf{x}}$ denotes the reconstructed signal (posterior mean of $\mathbf{x}$).

The corresponding qualities of the reconstructions of the two types of signal (uniform and non-uniform spikes) versus the number of measurements $K$ is summarized in Table 1 for the two algorithms. It is seen that if the number of measurements $K$ is larger than some specific values (140 (uniform spikes) or 100 (non-uniform spikes) for the Bottom-up algorithm, and 130 (uniform spikes) or 120 (non-uniform spikes) for the Top-down algorithm), the sparse signal can be exactly reconstructed. In this case, either there is only a single maximum in the evidence over $\boldsymbol{\alpha}$, or only a few insignificant local maxima occur, which do not affect the convergence of the algorithm to the global maximum.

If the number of measurements $K$ becomes smaller, the reconstructed results become unstable: many of the optimization runs produce the correct signal size of 20, and all of these reconstructions are quite close to a global maximum of the log evidence. However, the other runs give only local maxima of the evidence that correspond to larger amounts of non-zero signal components and large reconstruction errors. There are visible differences in the log evidence between the correctly reconstructed signal models and the incorrect ones. Fewer measurements result in larger average reconstruction errors and smaller differences in evidence between correctly reconstructed signal models and incorrectly reconstructed signal models.

**Table 1**
Quality of the signal reconstruction versus number of measurements *K* for an original signal length of *N*=512

| Quality of the reconstruction | Algorithm | Uniform spikes | Non-uniform spikes |
|---|---|---|---|
| All exact | Bottom-up | 140-512 | 100-512 |
|  | Top-down | 130-512 | 120-512 |
| Some unstable | Bottom-up | 60-140 | 40-100 |
|  | Top-down | 50-130 | 40-120 |
| All poor | Bottom-up | <60 | <40 |
|  | Top-down | <50 | <40 |

From these results, it is also found that the global maximum of the evidence corresponds to the most stable signal model; as shown on the right-hand side in Figure 1, one can observe that the spread is a lot more concentrated for the group of samples with reconstructed signal model size of around 20. The signal models with local maxima of the evidence are much less stable; they are very likely to converge to other signal models if they are perturbed slightly and used to restart the reconstruction algorithm.

The results clearly illustrate that the reconstruction process has poor robustness for the smaller values of *K* relative to *N*, the size of the original signal. Therefore, the specific values in Table 1 are the minimum number of measurements required for faithful reconstruction for this case, corresponding to a compression ratio (CR) for these two algorithms of $N/K = 3.66$ (uniform spikes) and 5.12 (non-uniform spikes), and 3.94 (uniform spikes) and 4.27 (non-uniform spikes), respectively.

However, if the number of measurements is smaller than 60 (uniform spikes) or 40 (non-uniform spikes) for the Bottom-up algorithm, and 50 (uniform spikes) or 40 (non-uniform spikes) for the Top-down algorithm, the difference in evidence between correctly reconstructed signal models and incorrect ones is negligible, which means that correct reconstruction through evidence maximization is not feasible.

Regarding the comparison of reconstruction errors between the cases of uniform and non-uniform spikes in Figures 1 and 2, the reconstruction errors of the local maxima for the non-uniform spikes tend to be smaller for the same *K* value. This observation is consistent with some theoretical analysis in Wipf and Rao (2006) which led to the conclusion that "the most difficult sparse inverse problem may involve weights of the same or even identical scale".

A comparison of the performance of the two algorithms shows that the reconstruction errors of the local maxima signal models and the minimum required number of measurements for faithful reconstruction are quite similar, which demonstrates that they have similar robustness for signal reconstruction. However, we are primarily interested in improving the performance of the Bottom-up algorithm in this paper because of its much greater efficiency for on-line applications to structural health monitoring.

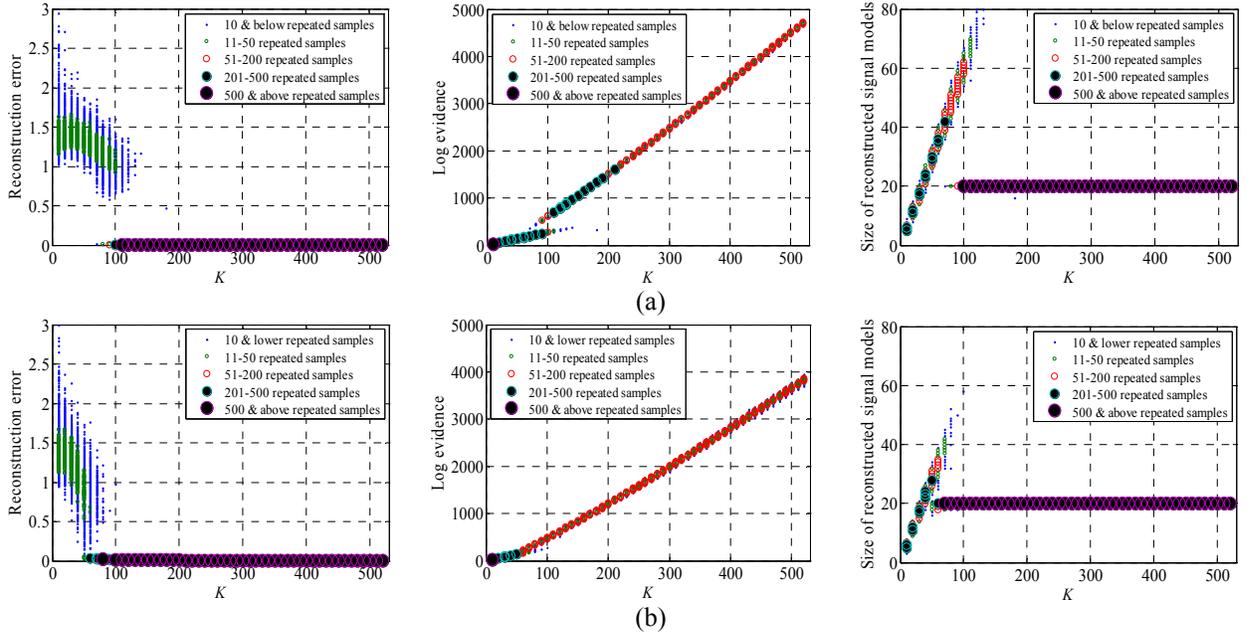

**Figure 1** The effects of number of measurements on the evidence and size of reconstructed signal models for the noise-free case run 1000 times (1000 samples) for each *K* using Bottom-up algorithm: (a) Uniform spikes; (b) Non-uniform spikes.

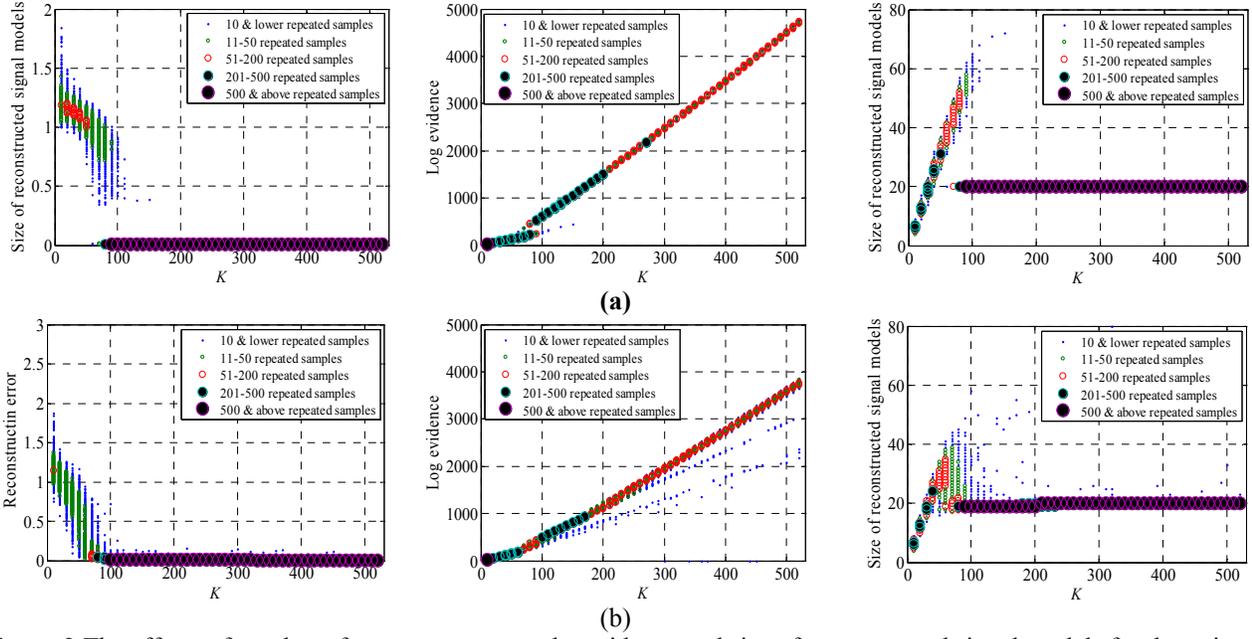

**Figure 2** The effects of number of measurements on the evidence and size of reconstructed signal models for the noise-free case run 1000 times (1000 samples) for each *K* using Top-down algorithm: (a) Uniform spikes; (b) Non-uniform spikes.

*Effect of measurement noise on reconstruction.* The effect of measurement noise (**r** in (2)) on the reconstruction is investigated in Figure 3 for the Bottom-up algorithm where the reconstruction error and log evidence are given as a function of the size of the reconstructed signal models for added zero-mean Gaussian noise with standard deviation at high levels of 5%, of the RMS of the synthetic measurements (compressive sensors will actually give lower noise to signal ratios). In these figures, we employ the same original signal and projection matrices as used in the noise-free case in Figure 1. The figures show that when the noise level increases, the reconstruction error tends to increase and the log evidence tends to decrease for the "optimal" signal models (the cluster of models of smaller sizes), indicating that a poorer reconstruction is produced. Most optimization runs produce reconstructions with signal sizes larger than 20, since the sizes of the reconstructed signal models need to be larger to "explain" the very noisy data.

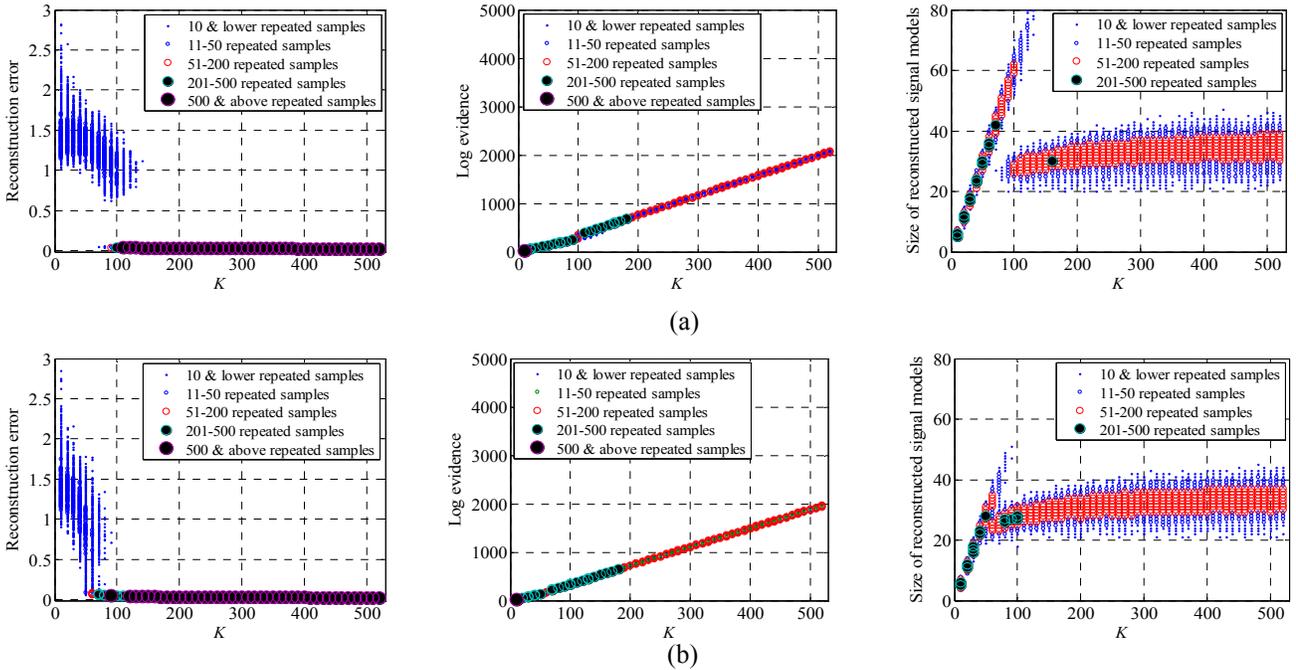

**Figure 3** The effects of number of measurements on the evidence and size of reconstructed signal models for the 5% noise case run 1000 times (1000 samples) for each *K* using Bottom-up algorithm: (a) Uniform spikes; (b) Non-uniform spikes.

*Effect of prediction error variance $\sigma^2$ on reconstruction.* In BCS reconstruction, the prediction error variance $\sigma^2$ is estimated along with the inverse prior variances for the coefficients **w**. However, the initial guess of the $\sigma^2$ value for the iterative scheme may affect the algorithm significantly due to the underdetermined nature of the inverse problem.

The variance $\sigma^2$ has significant influence on the trade-off between how well the reconstructed signal model fits the data and how sparse it is, which is related to the model complexity. There is an information-theoretical interpretation of the trade-off between data fitting and signal model complexity (Beck, 2010) that comes from expressing the log evidence in the following form:

$$\begin{aligned}
&\log[p(\mathbf{y}|\boldsymbol{\alpha}, \sigma^2)] \\
&= \int \log[p(\mathbf{y}|\mathbf{w}, \boldsymbol{\alpha}, \sigma^2)p(\mathbf{w}|\boldsymbol{\alpha}, \sigma^2)/p(\mathbf{w}|\mathbf{y}, \boldsymbol{\alpha}, \sigma^2)] \\
&\quad \times p(\mathbf{w}|\mathbf{y}, \boldsymbol{\alpha}, \sigma^2)d\mathbf{w} \\
&= \int \log[p(\mathbf{y}|\mathbf{w}, \boldsymbol{\alpha}, \sigma^2)]p(\mathbf{w}|\mathbf{y}, \boldsymbol{\alpha}, \sigma^2)d\mathbf{w} \qquad (23)\\
&\quad - \int \log[p(\mathbf{w}|\mathbf{y}, \boldsymbol{\alpha}, \sigma^2)/p(\mathbf{w}|\boldsymbol{\alpha}, \sigma^2)]p(\mathbf{w}|\mathbf{y}, \boldsymbol{\alpha}, \sigma^2)d\mathbf{w} \\
&= \mathbf{E}\big[\log\big(p(\mathbf{y}|\mathbf{w}, \boldsymbol{\alpha}, \sigma^2)\big)\big] \\
&\quad - \mathbf{E}\big[\log[p(\mathbf{w}|\mathbf{y}, \boldsymbol{\alpha}, \sigma^2)/p(\mathbf{w}|\boldsymbol{\alpha}, \sigma^2)]\big]
\end{aligned}$$

where $\mathbf{E}[\cdot]$ denotes the expectation with respect to the posterior PDF of **w**. The equation shows that log evidence is the difference between the posterior mean of the log likelihood function (the first term) and the relative entropy (or Kullback-Leibler information) between the prior and posterior distributions (the second term). The first term measures the average data-fit of the signal model $\mathcal{M}(\boldsymbol{\alpha}, \sigma^2)$, and the second term represents the information gain when updating $\mathcal{M}(\boldsymbol{\alpha}, \sigma^2)$ using the measurements **y**. It is the trade-off between these two terms that governs the evidence, and hence the posterior probability of $\mathcal{M}(\boldsymbol{\alpha}, \sigma^2)$.

The influence of $\sigma^2$ for Bayesian learning is investigated and shown in Figure 4 using synthetic data **y** with added measurement noise with standard deviation of RMS(**y**) × $10^{-5}$ where RMS(**y**) is the sample RMS of the data **y**. Figure 4(a) and (b) correspond to the uniform and non-uniform spikes, respectively, where the horizontal coordinate represents the relative value of $\sigma^2$ on a log scale. The vertical coordinate represents (i) reconstruction error;

(ii) number of non-zero terms in the reconstructed signal model; (iii) log evidence; (iv) data-fit measure; (v) the Kullback-Leibler information (the difference of (iv) and (v) gives the log evidence in (iii)). We run the Bottom-up Algorithm with one selected $\sigma^2$ and fix it for each case to obtain the posterior results. All the log evidence values, data-fit measure and the Kullback-Leibler information in Figures 4(a) (iii-v) and 4(b) (iii-v) are calculated with fixed $\sigma^2 = 0.01 \times \text{var}(\mathbf{y})$ so that the different optimal signal models for the different $\sigma^2$ values can be judged on a common basis. The original uniform and non-uniform spike signals are shown on the top of Figure 4.

As expected, the data-fit measure (the posterior mean of the log likelihood function) decreases with a lower specified prediction accuracy (larger $\sigma^2$), except for the case that the reconstructed signal model is optimal when there is a significant increase of the data-fit measure. Furthermore, it is observed that larger $\sigma^2$ produces smaller Kullback-Leibler information between the converged posterior PDF and the prior PDF, indicating that less information is extracted from the measurements by the reconstructed signal model. This is because larger $\sigma^2$ allows more of the data misfit to be treated as errors. Therefore, there are generally fewer terms in the signal models, as seen in Fig 4(a)(ii) and 4(b)(ii).

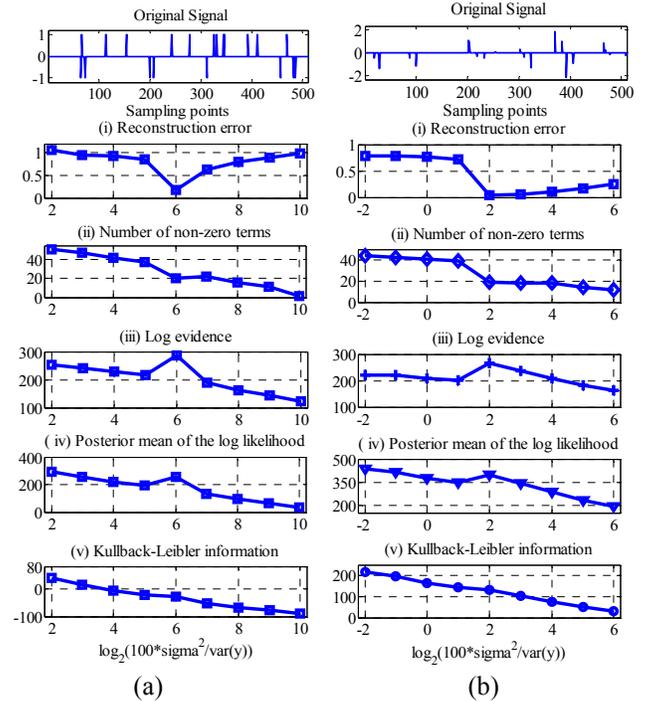

**Figure 4** The reconstructed signal model results using the Bottom-up algorithm with various prediction error variances $\sigma^2$. (a) Uniform spikes; (b) Non-uniform spikes (Note the different ranges on the y-axis between (a) and (b) for cases (iv) and (v)).

The most important point is that there is a peak of the log evidence profile in Figure 4(a) (iii) around $\log_2[100 \times \sigma^2/\text{var}(\mathbf{y})] = 6$, that is, $\sigma^2 = 0.64 \times \text{var}(\mathbf{y})$, which agrees with the lowest average reconstruction error; the corresponding reconstructed signal has the optimal trade-off between data-fitting and model complexity. For the non-optimal case, smaller $\sigma^2$ will produce an under-sparse signal model and larger $\sigma^2$ corresponds to an over-sparse signal. For non-uniform spikes, the optimal $\sigma^2$ of the evidence profile is at the point $\log_2[100 \times \sigma^2/\text{var}(\mathbf{y})] = 2$, that is, $\sigma^2 = 0.04 \times \text{var}(\mathbf{y})$.

## 3 PROPOSED BCS RECONSTRUCTION METHOD

The advantage of the Bottom-up Algorithm is its outstanding efficiency compared to the Top-down Algorithm. However, as we have illustrated, the Bottom-up Algorithm has a robustness problem. Therefore, an improved algorithm is presented to obtain a more robust performance for BCS reconstruction but with a reasonable tradeoff against computational effort.

### 3.1 Stochastic optimization over the prior variances

In the Bottom-up algorithm of sparse Bayesian learning, adding, deleting or re-estimating a basis vector for each iteration of the optimization over the individual hyperparameters (inverse prior variances) is based on whatever gives the maximum evidence increase. This is a deterministic optimization procedure. In this fast algorithm, the evidence change $\Delta\mathcal{L}_n$ for the $n^{th}$ term in (3) is controlled by the 'sparsity factor' $S_n$ and 'quality factor' $Q_n$ that are given in (18) and (19) (Tipping and Faul, 2003). Sparsity factor $S_n$ measures the extent that basis vector $\mathbf{\Theta}_n$ 'overlaps' the other basis vectors. The 'quality factor' $Q_n$ can be transformed to:

$$Q_n = \sigma^2 \mathbf{\Theta}_n^T (\mathbf{y} - (\mathbf{\Theta}\mathbf{w})_{-n}) \qquad (24)$$

It measures how well $\mathbf{\Theta}_n$ increases the evidence by helping to explain the data (Tipping and Faul, 2003). Both of these factors depend on the specific measurement information.

It is found that the value of the evidence change $\Delta\mathcal{L}$ for each iteration is sensitive to the details of the specific choice of projection matrix and the corresponding measurements when the number of measurements is smaller than a specific threshold (the lower limit in the first row of Table 1). The fewer the number of measurements, the more sensitive the evidence is to the details. This is because the evidence function has multiple significant local maxima when the ratio $K/N$ falls below a specific threshold and there exists no action (add, delete or re-estimate) that leads to an exceptionally large evidence increase $\Delta\mathcal{L}$ among the set of $\Delta\mathcal{L}$ for all possible actions. Thus, each time an action is chosen to maximize $\Delta\mathcal{L}$ in each iterative step of BCS, the choice is very sensitive to slight changes in the problem setting, such as the choice of projection matrix. Therefore, this deterministic search algorithm shows low robustness for finding the global maximum of the evidence over the hyperparameters, often finding sub-optimal local maxima rather than the optimal signal reconstruction.

This problem of getting stuck in local optima of the evidence function motivates trying a stochastic search method for optimization over the variances of the ARD prior, by introducing uncertainty into the search direction. One of the roles of the injected uncertainty in the proposed stochastic optimization process is to allow for a broader search of potentially unexplored areas which may contain the global maximum solution. Therefore, this approach significantly increases the probability of finding this global maximum, even when the evidence function has many local maxima.

Consider the two cases for the optimal $\alpha_n$ in (20), when $Q_n^2 > S_n$, $\frac{\partial^2 \mathcal{L}}{\partial \alpha_n^2}\big|_{\alpha_n = \frac{S_n^2}{Q_n^2 - S_n}} = \frac{-S_n^2}{\alpha_n^2(\alpha_n + S_n)^2}$ is always negative, therefore $\mathcal{L}(\alpha_n)$ has a unique maximum at $\alpha_n = \frac{S_n^2}{Q_n^2 - S_n}$. For the case of $Q_n^2 < S_n$, the maximum point of the $\mathcal{L}(\alpha_n)$ is at $\alpha_n = \infty$ (Tipping and Faul, 2003). This means that the successive action of adding, deleting or re-estimating the individual basis vector by analytical updating $\alpha_n$ in (20) will always increase the log evidence $\mathcal{L}(\alpha_n)$, and the corresponding change of log evidence $\Delta\mathcal{L}_n$ is positive. The basis vector $\mathbf{\Theta}_n$ with larger $\Delta\mathcal{L}_n$ should have larger probability to be updated when searching for the global maximum of $\mathcal{L}(\boldsymbol{\alpha})$, so we set the probability of accepting the action of addition, deletion or re-estimation of the candidate $n^{th}$ basis vector as

$$p_n = \Delta\mathcal{L}_n / \max(\Delta\mathcal{L}_n) \qquad (25)$$

where $\Delta\mathcal{L}_n$ is the change of log evidence produced by the $n^{th}$ basis vector being added, deleted or re-estimated and $\max(\Delta\mathcal{L}_n)$ is the largest increase of the log evidence in a specific step. When the $n^{th}$ basis vector is not included in the current signal model and $Q_n^2 < S_n$, we consider $p_n = 0$. Thus, if the evidence change $\Delta\mathcal{L}_n$ is the largest among all basis vectors ($p_n = 1$), then the action of this candidate basis vector is definitely accepted. If the evidence change $\Delta\mathcal{L}_n$ is not the largest ($p_n < 1$), then the action of the candidate $n^{th}$ basis vector is accepted with probability $p_n$. We can accomplish this process by generating an independent value $u_n$ from the uniform distribution $U(0,1)$. If $p_n \geq u_n$, the corresponding action of the $n^{th}$ basis vector is implemented, otherwise it is skipped. Finally, we use the same termination criterion as the Bottom-up Algorithm, which judges the convergence by the change in all $\log \alpha_n's$ being sufficiently small.

It is noted that the stochastic optimization may produce signal models with size $N' > K$, because it may add multiple terms in one iteration. In this case, we switch to the Top-down algorithm to prune out a number of terms.

The reason for doing this is that the Top-down algorithm tries to increase the evidence at each iteration by updating all the basis vector terms of the current signal model while the Bottom-up algorithm successively optimizes individual ones in each iteration. The matrix inversion involves $O(K^3)$ operations for each iteration of the two algorithms.

## 3.2 Optimization over all hyperparameters by successive relaxation

In the previous subsection, we proposed using stochastic optimization to maximize the evidence over the inverse prior variances. The remaining issue is how to optimize the evidence with respect to the prediction error variance $\sigma^2$. We have found that some care is needed because of the important influence of $\sigma^2$ on the evidence function but that successive relaxation works well where we optimize the evidence in the full hyperparameter space $[\alpha, \sigma^2]$ in an iterative fashion: first $\alpha$ is optimized with $\sigma^2$ fixed and then $\sigma^2$ is optimized with $\alpha$ fixed at its intermediate optimal value, this procedure being repeated until convergence is achieved.

To initialize the algorithm, we select the single inverse prior variance of the $n^{th}$ term that maximizes $\|\Theta_n^T \mathbf{y}\|^2 / \|\Theta_n\|^2$ over $n$ as $\alpha_n = 1$. All other inverse prior variances are set to infinity, implying that the corresponding terms are excluded in the initial signal model. We update the prediction error variance $\sigma^2$ based on this signal model, then fix this updated $\sigma^2$ and optimize the intermediate evidence function to obtain a set of improved optimal inverse prior variances. Because the updated $\sigma^2$ is optimal for maximizing the evidence function based on the current signal model, there is a high probability for the intermediate optimal hyperparameters α to be near the global maximum. Furthermore, the stochastic optimization procedure introduced in Section 3.1 allows a non-zero probability to accept candidates with lower evidence that helps the algorithm to escape being trapped in local maxima.

At each intermediate step, once the intermediate optimum inverse prior variances are found, the updated $\sigma^2$ for the next iteration is estimated based on (14) with $a = 1$:

$$(\sigma^2)^{[j+1]} = \frac{\|\mathbf{y} - \Theta \mathbf{\mu}^{[j]}\|^2 + 2b^{[j]}}{K - \sum_{n=1}^{N_j}(1 - \alpha_n^{[j]} \Sigma_{nn}^{[j]})} \quad j = 1,2,3 \ldots J - 1$$

where $N_j$ is the current size of the signal model; and $\mathbf{\mu}^{[j]}$ and $\Sigma^{[j]}$ are the posterior mean and covariance of the multivariate Gaussian distribution over the reconstructed sparse coefficients $\mathbf{w}$ from the $j^{th}$ iteration, calculated by (9) and (10), respectively. We start with $b^{[1]} = 0$, then parameter $b^{[j+1]}$ can be updated from Eq (15), using the current estimated parameter values at each iteration.

## 3.3 BCS reconstruction algorithm using stochastic optimization and successive relaxation

The ideas in the previous two subsections are combined to produce an improved BCS reconstruction method that iterates between inner and outer loop optimizations. The outer loop updates the prediction-error variance $\sigma^2$ and is terminated when the changes in the reconstructed signal models are sufficiently small, e.g. $\|(\hat{\mathbf{x}})^{[j+1]} - (\hat{\mathbf{x}})^{[j]}\|_2^2 / \|(\hat{\mathbf{x}})^{[j]}\|_2^2 < \epsilon$, a specified threshold (we take $\epsilon =$ measurement noise level in the examples later). The inner loop performs the stochastic optimization procedure over the prior variances $\alpha$ and is terminated when the change in all $\log \alpha_n's$ is less than $10^{-6}$. We present two versions of the improved BCS algorithms: BCS-SO with the prior parameters for $\sigma^2$ fixed at $a = 1, b = 0$; and BCS-SO* which uses equation (15) for the optimal $b$ with $a = 1$ (optimization of parameter $a$ from (16) does not converge to a finite value, so we fix $a = 1$).

| Algorithms BCS-SO and BCS-SO*: Improved BCS with stochastic optimization and successive relaxation |
|---|
| 1. Inputs: $\Theta$, $\mathbf{y}$; Outputs: posterior mean and covariance of $\mathbf{w}$ and $\mathbf{x}$ |
| 2. Initialize all inverse prior variances to a very large value except set $\alpha_n = 1$ for the term that maximizes $\|\Theta_n^T \mathbf{y}\|^2 / \|\Theta_n\|^2$, then initialize $\sigma^2$ using (26) and set parameter $b^{[1]} = 0$. |
| 3. **While** convergence criterion on the $\log \alpha_n's$ is not met (Inner loop) |
| 4. Calculate $p_n$ in (25) for every basis vector $\Theta_n$ and update $\Theta_n$ if $p_n \geq u_n$, where $u_n$ is sampled from $U(0,1)$. Calculate $S_n$ and $Q_n$ from (19) and (20). For updating $\Theta_n$ for all $n$, choose a random order covering all values. |
| 5. **If** $Q_n^2 - S_n > 0$ and $\alpha_n = \infty$, add $\Theta_n$ and update $\alpha_n$ using (21) |
| 6. **If** $Q_n^2 - S_n > 0$ and $\alpha_n < \infty$, re-estimate $\alpha_n$ using (21) |
| 7. **If** $Q_n^2 - S_n \leq 0$ and $\alpha_n < \infty$, delete $\Theta_n$ and set $\alpha_n = \infty$ |
| 8. **End if** |
| 9. Update $\mathbf{\mu}$ and $\Sigma$ in (9) and (10). |
| 10. Check if the number of non-zero terms $N'$ of the current signal model is larger than the number of measurements $K$; if not, go to step 11. Otherwise, employ the Top-down algorithm to update $\alpha$ using (13) until the termination criterion on $\log \alpha_n's$ is met, then update $\sigma^2$ using (26) and end algorithm. For initializing Top-down algorithm, the current signal model is expanded to a full signal model by setting the prior variances for terms not included in the current signal model equal to the average prior variance of the current signal coefficients. |
| 11. **End while** (the intermediate optimal inverse prior variances are then established for the current $\sigma^2$) |

12. Update $\sigma^2$ using (26) based on the current inverse prior variances. If BCS-SO, set all $b^{[j]} = 0$, and if BCS-SO*, set $b^{[j+1]} = (\sigma^2)^{[j+1]}$.
13. Check if updating of reconstructed signal model $\hat{\mathbf{x}} = \mathbf{\Psi\mu}$, has converged. If not, use the current intermediate optimal hyperparameters ($\boldsymbol{\alpha}$ and $\sigma^2$) as initial information and repeat the inner loop (steps 3 to 11); otherwise end.

At step 4 of the algorithm, the stochastic optimization scheme is employed and the candidate basis vector $\boldsymbol{\Theta}_n$ for updating is selected with probability $p_n$ calculated by (25). Another important step in the algorithm is the updating of the prediction error variance $\sigma^2$ based on the current converged inverse prior variances $\boldsymbol{\alpha}$, which uses the idea of successive relaxation as shown in step 12. Finally, updating of the posterior mean and covariance of **w** and **x** is done as in the original BCS reconstruction algorithm.

We note that the proposed stochastic optimization algorithm is less efficient than the original deterministic Bottom-up optimization algorithm. There is a fundamental tradeoff between algorithmic efficiency and robustness (reliability and stability) in the optimization procedure.

## 4 EXAMPLE RESULTS

### 4.1 Synthetic sparse spike signals

We compare results of our modified algorithms (BCS-SO and BCS-SO*) in Section 3.3 with corresponding results of several other published algorithms used for reconstruction of CS signals. We denote the original Bottom-up algorithm (Ji et al., 2008; Tipping and Faul, 2003) with a constant prediction error variance $\sigma^2 = \text{var}[\mathbf{y}] \times 0.1$ as BCS-B-F, and the Bottom-up algorithm (Tipping and Faul, 2003) which updates $\sigma^2$ every 5 iterations using (14) with $a=1$, $b=0$ as BCS-B-U. The Matlab code for BCS-B-F was downloaded from http://people.ee.duke.edu/~lcarin/BCS.html while the code for optimization of the hyperparameters $\boldsymbol{\alpha}$ and $\sigma^2$ for BCS-B-U is from http://www.vectoranomaly.com/SparseBayes.htm.

The original Top-down algorithm (Tipping, 2001), denoted BCS-T, is also introduced using the codes from http://www.miketipping.com/sparsebayes.htm; it updates $\sigma^2$ in each iteration using (14) with $a=1$, $b=0$. We also make a comparison to the BP algorithm (Candes et al., 2006), which is a norm-minimization algorithm that involves linear programming, by using the $l1$-magic package at http://www-stat.stanford.edu/~candes/l1magic/.

We consider signals which are the same as in Figure 1: one is uniform $\pm 1$ random spikes (Figure 5(a)(i) and 5(b)(i)), and the other is zero-mean unit variance Gaussian spikes (Figure 5(c)(i) and 5(d)(i))). We show one result of optimal signal reconstruction for BCS-SO and BCS-SO* (Figure 5(a) (vi-vii) and 5(c) (vi-vii)))) and one suboptimal result (Figure 5(b) (vi-vii) and 5(d) (vi-vii)). Figures 5(a)(ii), 5(b)(ii), 5(c)(ii) and 5(d)(ii)) demonstrate reconstruction results for the two types of signals using the original BP algorithm while Figure 5(a)(iii-vii), 5(b)(iii-vii)), 5(c)(iii-vii) and 5(d)(iii-vii)) show the results of five BCS algorithms. Because of insufficient number of measurements, sub-optimal signal representations are obtained for BCS-B-F, BCS-B-U and BCS-T in all results. However, BCS-SO and BCS-SO* produces almost perfect reconstructions in Figures 5(a)(vi-vii) and 5(c)(vi-vii) for uniform and non-uniform cases with a compression ratio $CR$ of $N/K$ =5.69 and 8.53, respectively.

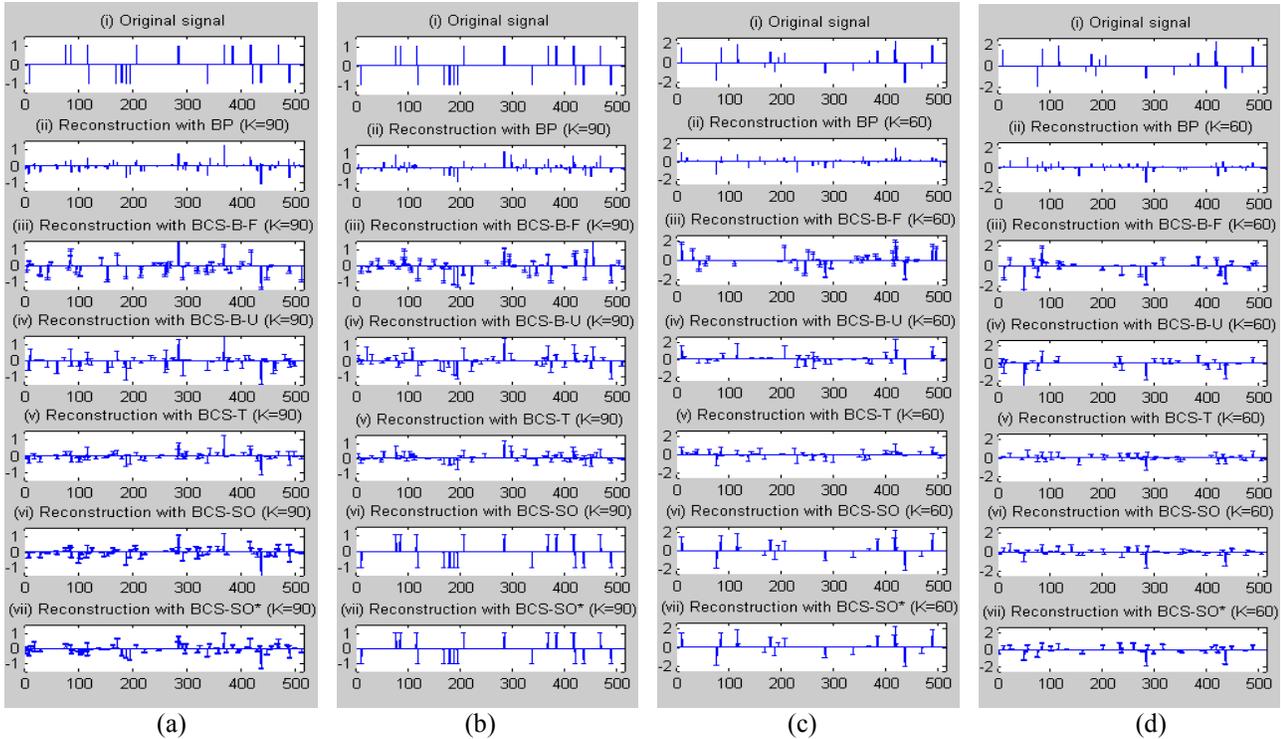

**Figure 5** Original and reconstructed signals of length N=512 using 6 different algorithms for Uniform spikes: (a) Optimal and (b) Sub-optimal cases; and Non-uniform spikes: (c) Optimal and (d) Sub-optimal cases. The short horizontal lines give the error bars on the reconstructed signal for the Bayesian algorithms.

The error-bars (defined as $\pm$ one standard deviation) for the reconstruction $\mathbf{x}$ that are produced by the posterior uncertainty in $\mathbf{w}$, are also shown in the results; they are computed from the diagonal elements of the covariance matrix of $\mathbf{x}$, $\mathbf{\Psi\Sigma\Psi^T} = \mathbf{\Sigma}$ (since $\mathbf{\Psi} = \mathbf{I_N}$), where $\mathbf{\Sigma}$ is given by (10) for the optimal hyperparameters. Since the BP algorithm is a norm-minimization algorithm, it does not quantify the uncertainty in its reconstructed signals. For the BCS methods, the error bars estimated for the suboptimal signal model (Figures 5(b)(vi-vii) and 5(d)(vi-vii)) by BCS-SO and BCS-SO* are larger compared with those estimated for the optimal results (Figures 5(a)(vi-vii) and 5(c)(vi-vii)), which indicates that the signal reconstruction confidence is smaller, even though the uncertainty is underestimated because the algorithm has given a local maximum of the evidence (instead of the global maximum), so the MAP approximation in (11) is poor. However, for the original BCS methods (BCS-B-F, BCS-B-U and BCS-T), the estimated posterior error bars generally cannot be used for judging signal reconstruction confidence, which will also be shown in later results.

In Figures 6, the results for the reconstruction error and average estimated error bars for 50 trials with different projection matrices but the same original signal are shown for BCS-SO and BCS-SO*. The number of measurement is 90 and 60 for uniform and non-uniform spikes signal, respectively, and three noise levels of 0.001%, 2% and 5% are compared in these results. The average error bars are computed as the mean of diagonal elements of the covariance matrix of $\mathbf{x}$, $\mathbf{\Psi\Sigma\Psi^T} = \mathbf{\Sigma}$ (since $\mathbf{\Psi} = \mathbf{I_N}$). It can be observed that better uncertainty estimation performance is obtained for the BCS-SO*. It is seen that for BCS-SO* when the measurement noise level is 0.001% (Figure 6 (b)(i) and 6 (d)(i)), all suboptimal reconstructed signal models give average error bars that are much larger than others, while for the optimal signal models they are relatively small. This correspondence can also be observed for BCS-SO ((Figure 6 (a)(i) and 6 (c)(i))), although it is not so obvious as BCS-SO*. Therefore, the error bars can be considered as a useful tool for signal reconstruction diagnostics. However, for heavier noise cases of 2% and 5%, (Figure 6 (a)(ii-iii), 6 (b)(ii-iii), 6 (c)(ii-iii) and 6 (d)(ii-iii)), there are only a few average error bars that are large and correspond to incorrect signal reconstructions. One of the reasons for these results is that the $\epsilon$ value in the termination criterion $\left\|(\hat{\mathbf{x}})^{[j+1]} - (\hat{\mathbf{x}})^{[j]}\right\|_2^2/(\hat{\mathbf{x}})^{[j]} < \epsilon$ for BCS-SO and BCS-SO* is increased for higher noise cases to avoid excessive numbers of iterations.

In the next examples, the performance is investigated for different compression ratios $CR$. We fix the signal length as $N = 512$ and the number of non-zero coefficients $T = 20$, and vary $K$ from 40 to 120 (compression ratios from 4.27 to 12.8). A Gaussian random projection matrix is constructed for each experiment and the associated reconstruction errors are calculated as $\|\hat{\mathbf{x}} - \mathbf{x}\|_2^2/\|\mathbf{x}\|_2^2$, where $\hat{\mathbf{x}}$ and $\mathbf{x}$ are the reconstructed and original signal vectors, respectively. Because of the randomness of the projection matrix for the CS measurements in this set of experiments, we execute the experiment 100 times and report the average performance.

To investigate how the compression ratio $CR$ affects the reconstruction performance, we compare the six algorithms: BP, BCS-B-F, BCS-B-U, BCS-T, BCS-SO and BCS-SO* for different compression ratios. We repeat the same experiment with two different noise levels of 0.001% and 5%, and the results are shown in Figure 7. In these results, different thresholds of acceptable reconstruction error (0.01, 0.10 and 0.50) are employed to investigate the corresponding rates of acceptable performance. As expected, increases in $CR$ correspond to a decrease in rates of acceptable performance for these methods. BCS-SO* has a little better signal reconstruction performance (higher rate) than BCS-SO. For the nearly noise-free case (noise level=0.001%) shown in Figure 7 (i), BCS-SO and BCS-SO* clearly outperform all other methods, especially when the reconstruction error threshold is low (0.01). The critical value of compression ratio below which nearly perfect lossless reconstruction performance (RE<0.01) occurs is seen to be substantially larger for the BCS-SO and BCS-SO* algorithms than for the other three BCS algorithms (a difference of about 1.0 in the critical compression ratio values for BCS-B-F and BCS-B-U). In Figure 7 (ii) with noise level of 5%, the rates of acceptable performance for higher reconstruction error thresholds (0.1 and 0.5) are slightly smaller than that in the 0.001% noise level case (Fig 7 (i)), as expected, while there is no $CR > 4$ that gives a reconstruction error below the threshold of 0.01 for any of the six algorithms. The BCS-SO and BCS-SO* algorithms show strong robustness to noise. The superior performance of BCB-SO and BCS-SO* is demonstrated by its much higher rates of acceptable performance than that of BCS-B-F, BCB-B-U and BCS-T.

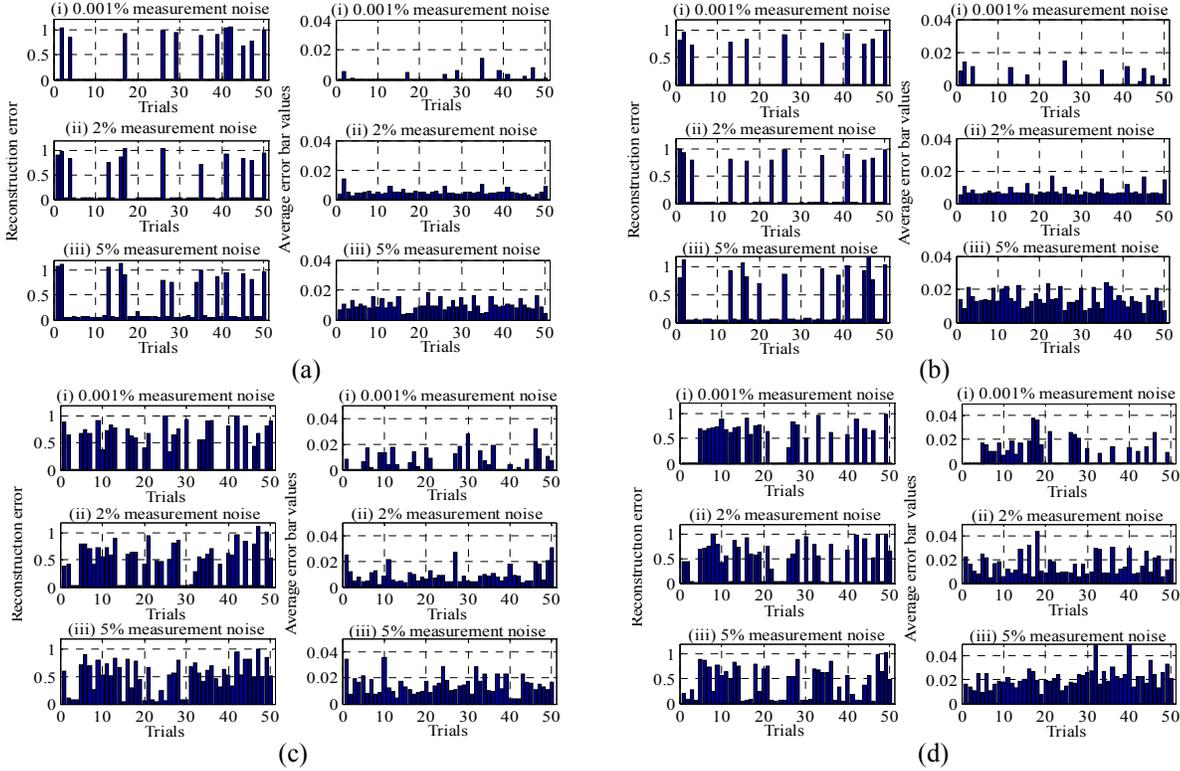

**Figure 6** Reconstruction errors and the corresponding average error bars over 50 runs for various noise levels: (a) BCS-SO, Uniform spikes, $K=90$; (b) BCS-SO*, Uniform spikes, $K=90$; (c) BCS-SO, Non-uniform spikes, $K=60$; (d) BCS-SO*, Non-uniform spikes, $K=60$.

From the comparison of performance of the two types of signals (uniform and non-uniform spikes), it is seen that the performance improvement using BCS-SO and BCS-SO* is more obvious for non-uniform spikes. Furthermore, it is observed that for a given compression ratio and reconstruction error threshold, the rates of acceptable performance for the non-uniform spikes is much higher than that for the uniform spikes, even for the cases with higher levels of contaminating noise. These results are consistent with the conclusion from Figures 1-3: higher diversity of the weight magnitudes gives a better chance of finding the optimal solution.

The corresponding computational times for signal reconstruction for uniform spikes and non-uniform spikes on a 2.67 GHz PC are given in Tables 2, respectively. It is seen that the BCS-B-F is the most efficient method, while the BCS-SO and BCS-SO* algorithms are less efficient compared with BCS-B-F, requiring about 6-18 times more computational effort. However, for BCS-SO and BCS-SO*, it can be inferred that signal reconstruction using this PC, if good reconstruction is consistently achieved, can keep up for on-line operation when the sampling frequency of the compressive sensors is less than 350Hz.

**4.2 Application to bridge accelerometer data**

Tianjin Yonghe Bridge (Figure 8) is one of the earliest cable-stayed bridges constructed in the mainland of China. It has a total length of 512 m, comprising a main span of 260m and two side spans of 125m. A sophisticated long-term structural health monitoring system was designed and implemented on this bridge. The system includes optical fiber Bragg-grating (FBG) strain sensors, accelerometers, electromagnetic sensors, GPSs, anemometer and temperature sensors. Fourteen uniaxial accelerometers and one biaxial accelerometer were permanently installed on the deck of the main span and two side spans, and on one tower top.

**Table 2**
Average computational time in seconds of six algorithms for different noise levels corresponding to the results in Figures 5-7 with original signals of uniform and non-uniform spikes.

| signal | Noise level | BP | BCS-B-F | BCS-B-U | BCS-T | BCS-SO* | BCS-SO |
|---|---|---|---|---|---|---|---|
| Uniform spikes | 0.001% | 0.269 | 0.118 | 0.124 | 0.501 | 1.320 | 1.364 |
|  | 5% | 0.313 | 0.135 | 0.232 | 0.736 | 1.167 | 1.251 |
| Non-Uniform spikes | 0.001% | 0.292 | 0.071 | 0.082 | 0.618 | 1.178 | 1.252 |
|  | 5% | 0.356 | 0.086 | 0.232 | 0.816 | 0.821 | 0.911 |

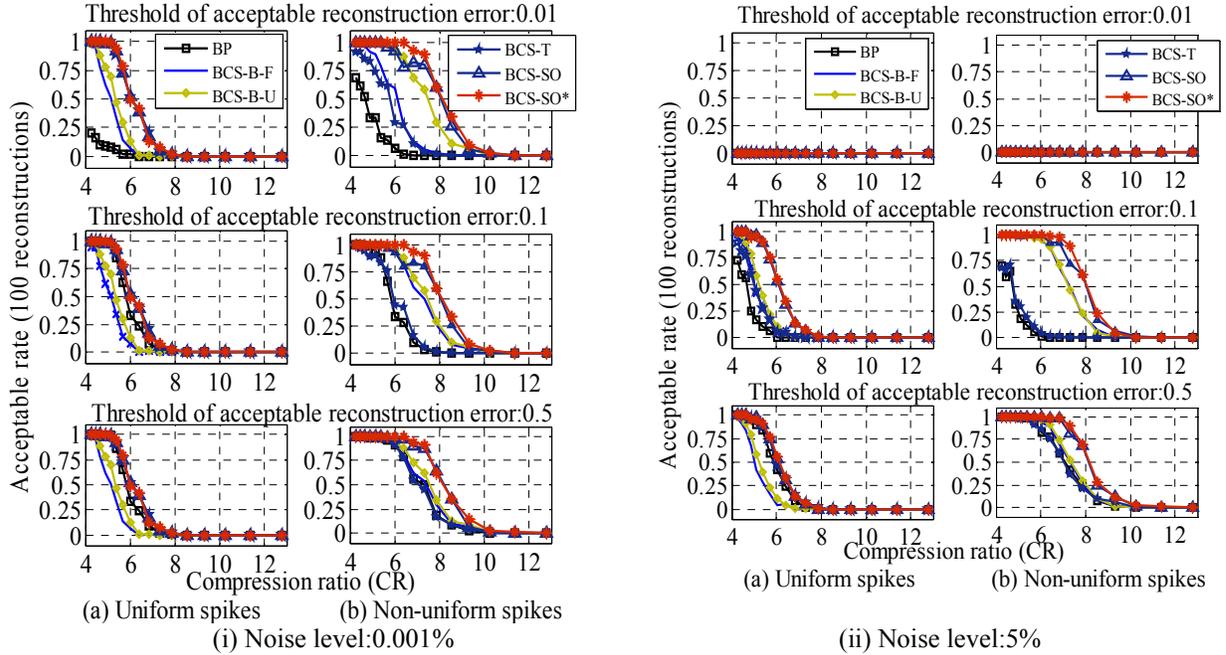

**Figure 7** Relation of compression ratio and the rate of acceptable reconstruction error for 3 different thresholds of the latter: (a) Uniform spikes; (b) Non-uniform spikes; for two different noise levels and six reconstruction algorithms

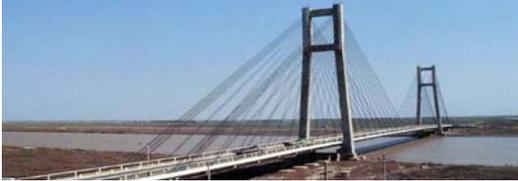

**Figure 8** Photo of the Tianjin Yonghe Bridge

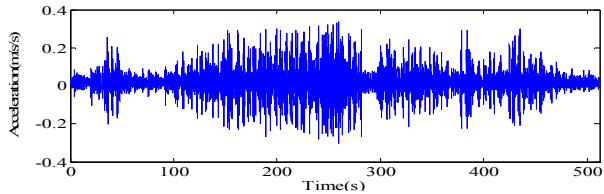

**Figure 9** Acceleration response of bridge deck from an accelerometer at mid-span.

The signal from one accelerometer which is installed on the deck of the main span is employed here. Figure 9 shows the acceleration response time history of length 512 seconds from this accelerometer. The signal has 51200 samples at a sample frequency of 100 Hz.

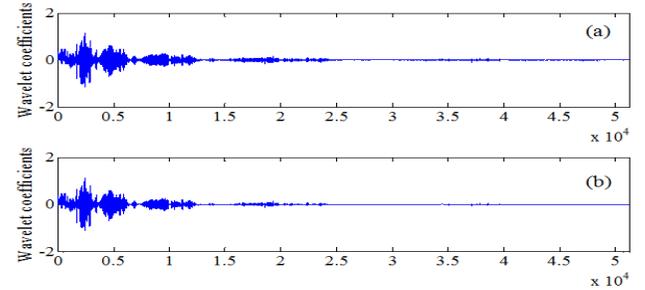

**Figure 10** Wavelet coefficients of the acceleration data using the dB1 wavelet basis: (a) original wavelet coefficients; (b) de-noised wavelet coefficients.

Using the dB1 wavelet basis, the wavelet coefficients of the whole acceleration data are computed and shown in Figure 10(a), where it is seen that only a small number ($m$) of the $N$=51200 wavelet coefficients are significant, and the other ($N-m$) coefficients are small. After using hard threshold de-noising, so that all coefficients of magnitude smaller than $10^{-4}$ are set to be zero, the wavelet coefficients are sparse in the dB1 wavelet basis domain because only $m=26026$ wavelet coefficients are nonzero, as shown in Figure 10(b).

We investigate the performance of BCS for vibration signals used in SHM by dividing the signal in Figure 9 into 100 segments of $N=512$ and compressing each segment **x** by projection using the same random projection matrix for each segment, as in (2): $\mathbf{y}=\mathbf{\Phi x}=\mathbf{\Theta w},$ where $\mathbf{\Theta}=\mathbf{\Phi\Psi}$ and $\mathbf{\Psi}$ is the **dB1** wavelet basis matrix and $\mathbf{w}=\mathbf{\Psi}^{-1}\mathbf{x} = \mathbf{\Psi}^T\mathbf{x}$ is known. In practice, one would acquire data already in a compressed form from CS accelerometers, where the projection arithmetic is integrated with the analog-to-digital converter into the sensor itself, so $\mathbf{x}$ and $\mathbf{w}$ are unknown.

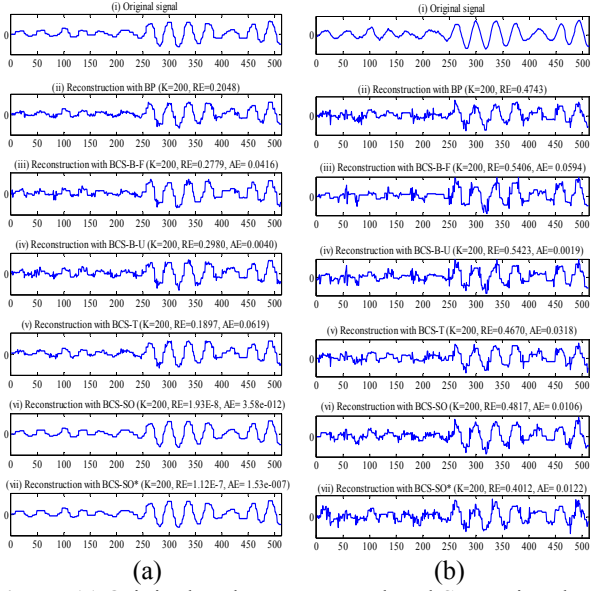

**Figure 11** Original and reconstructed real SHM signals of length N=512: (a) Case 2; (b) Case 1. (RE and AE denotes corresponding reconstruction error and average estimated error bars for all nonzero terms, respectively.)

The 100 segments, each of original length $N$=512, allow us to investigate the BCS reconstruction performance for an SHM signal which is not sparse but is approximately so. We also consider a second case in order to investigate the performance of the BCS reconstruction algorithms for a sparse signal. We invert the de-noised wavelet coefficients shown in Figure 10(b) and divide the time-domain signal into 100 segments of $N = 512$. Each segment $\mathbf{x}_d$ is then compressed using the same projection matrix for each segment. In this case, measurements vector $\mathbf{y}_d$ is expressed as

$$\mathbf{y}_d=\mathbf{\Phi}\mathbf{x}_d=\mathbf{\Theta}\mathbf{w}_d \qquad (27)$$

where $\mathbf{w}_d = \mathbf{\Psi}^{-1}\mathbf{x}_d = \mathbf{\Psi}^T\mathbf{x}_d$ is known. We then obtain optimal reconstructed coefficients $\widehat{\mathbf{w}}$ and $\widehat{\mathbf{w}}_d$ from measurements $\mathbf{y}$ and $\mathbf{y}_d$, respectively, using the Bayesian reconstruction algorithms. These optimal vectors are the posterior means given by (9) and (10) using the optimal hyperparameters $\widehat{\boldsymbol{\alpha}}$ and $\widehat{\sigma}^2$, calculated separately for data $\mathbf{y}$ and $\mathbf{y}_d$. The corresponding optimal reconstructed accelerations $\widehat{\mathbf{x}}$ and $\widehat{\mathbf{x}}_d$ are then obtained by a wavelet transform using the reconstructed wavelet coefficients: $\widehat{\mathbf{x}} = \mathbf{\Psi}\widehat{\mathbf{w}}$ and $\widehat{\mathbf{x}}_d = \mathbf{\Psi}\widehat{\mathbf{w}}_d$. The CS reconstruction errors for the signals recovered from the measurements $\mathbf{y}$ and $\mathbf{y}_d$ are calculated based on the following expressions as: $R = \|\mathbf{x} - \widehat{\mathbf{x}}\|_2^2 / \|\mathbf{x}\|_2^2$ and $R_d = \|\mathbf{x}_d - \widehat{\mathbf{x}}_d\|_2^2 / \|\mathbf{x}_d\|_2^2$.

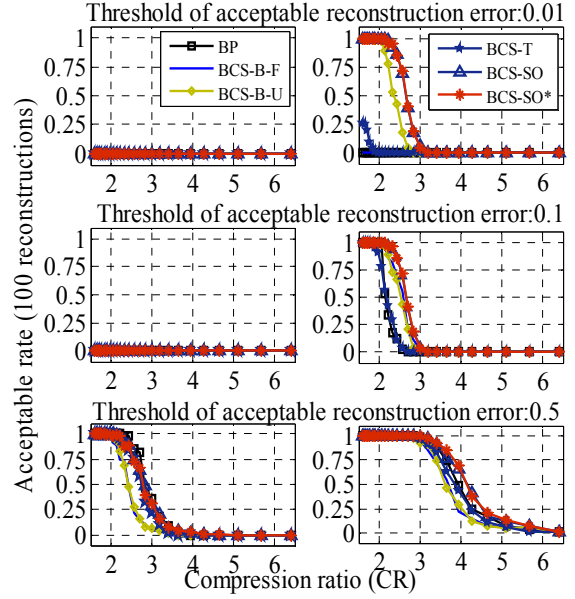

**Figure 12** Relation of compression ratio and the rate of acceptable reconstruction error for 3 different thresholds for six algorithms using real SHM acceleration data: (a) Case 1; (b) Case 2.

It is also of interest to compare the sparsity in the reconstructed coefficients $\widehat{\mathbf{w}}$ and the original de-noised coefficients $\mathbf{w}_d$, so we define a *sparsity ratio SR* by:

$$SR = \hat{s}/s_d \qquad (28)$$

where $\hat{s}$ is the total number of wavelet coefficients which are smaller than the threshold of $10^{-4}$ in vector $\widehat{\mathbf{w}}$, representing the sparsity of the reconstructed acceleration signal $\widehat{\mathbf{x}}$ with respect to the wavelet basis $\{\mathbf{\Psi}_n\}$; and $s_d$ is the sparsity of the de-noised wavelet coefficients vector $\mathbf{w}_d$. We also use the optimal reconstructed coefficient vector $\widehat{\mathbf{w}}_d$ to calculate a sparsity ratio $SR_d = \hat{s}_d/s_d$, where $\hat{s}_d$ is the sparsity of the reconstructed acceleration signal $\widehat{\mathbf{x}}_d$ with respect to the wavelet basis $\{\mathbf{\Psi}_n\}$.

Bao et al. (2011) investigated the norm-minimization algorithm BP for CS reconstruction by using real acceleration data collected from a SHM system. In this study, the BP algorithm (Candes et al., 2006) is also applied together with the BCS-B-F, BCS-B-U, BCS-T, BCS-SO and BCS-SO* algorithms to make a comparison of reconstruction errors and sparsity ratio as shown in Figures 12 and 13. In Figures 12 and 13, each evaluated point on the curves is computed based on the results for the 100 segments described before and the same random projection matrix is chosen for each reconstruction. We define the reconstructions corresponding to measurements $\mathbf{y}$ and $\mathbf{y}_d$ as Case 1 and Case 2, respectively. Case 1 is the real non-sparse case which should be tackled in structural health monitoring. The $\epsilon$ values in the termination criterion

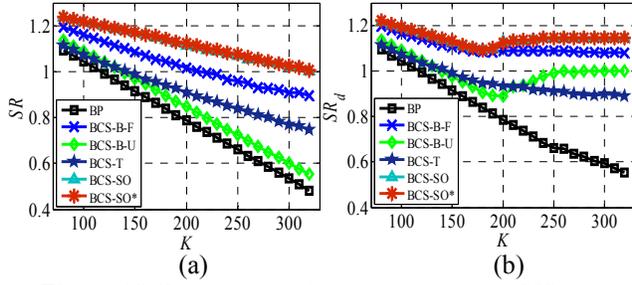

**Figure 13** Comparison of sparsity ratio over 100 runs of various algorithms as a function of *K* using real SHM acceleration data: (a) average results of six algorithms for Case 1 (non-sparse signals); (b) average results of six algorithms for Case 2 (sparse signals).

$\|(\hat{\mathbf{x}})^{[j+1]} - (\hat{\mathbf{x}})^{[j]}\|_2^2 / \|(\hat{\mathbf{x}})^{[j]}\|_2^2 < \epsilon$ of BCS-SO* is set to be 0.05 and 0.00001 for Case 1 and 2, respectively.

The plot of the posterior mean of reconstructed signals for each method when number of measurements *K*=200 ( *CR* =2.56) is shown in Figure 11 compared with the original signals for Case1 and Case 2. The corresponding reconstruction errors (RE) and average posterior error bars (AE) for these reconstructed signals are also given. Similar conclusions to those for synthetic data are obtained for Case 2: better reconstruction performance is obtained by BCS-SO and BCS-SO* relative to other methods. Furthermore, the error bars are larger when suboptimal solutions are obtained for the BCS-SO and BCS-SO* methods. For Case 1, the wavelet coefficient vector for the original signal is not so sparse and contains a lot of little spikes (Figure 10) which will be considered as modeling error for sparse signal reconstruction. It contaminates the performance of the proposed BCS-SO and BCS-SO* methods. There is no method that can produce perfect reconstruction in Case 1.

We also set different thresholds for acceptable reconstruction errors to examine the reconstruction performance for various methods. From the observation of the results on the right-hand side of Figure 12 for Case 2 (sparse signals), the proposed BCS-SO and BCS-SO* algorithms outperform all other methods, especially when the threshold is 0.01. The proposed BCS-SO and BCS-SO* algorithms can achieve exact lossless compression performance (RE<0.01) when the compression ratio is smaller than 2.2. For Case 1, it is seen that BP performs better than all Bayesian CS methods for the case of low compression ratio (2.25-2.75) when the threshold is 0.5**,** though the rates of acceptable performance for BP and BCS-SO* are close. There are no acceptable reconstructions when the thresholds are 0.01 and 0.1 for any of the six algorithms.

From the comparison of the sparsity ratio for both Cases 1 and 2, as shown in Figure 13, the BP algorithm produces less sparse signals than the de-noised signal ( $SR < 1$ ), which is also the reason why its rates of acceptable performance for the reconstruction error threshold of 0.01 and 0.1 are much lower (Figure 12). In contrast, the reconstructed coefficient vectors of the five BCS algorithms produce more sparse signals than BP and comparable or better sparsity than the de-noised signal. This is beneficial effect of the central feature of BCS methods, that the effective dimensionality of the signal model (equivalent to the number of retained coefficients) is determined automatically as part of the fully Bayesian inference procedure. Also, the algorithm can add and delete basis vectors analytically by Eq. (21). The increased numerical precision of the automatic operation is the likely explanation for the improvement in sparsity demonstrated by the BCS algorithms.

The average error bars and reconstruction errors over 50 trials of six algorithms for Cases 1 and 2 with the number of measurements *K*=200 are shown in Figure 14. The perfect correspondence of larger posterior error bars with suboptimal reconstruction can be observed for BCS-SO* in Case 2 (Figure 14(b)(vi)), although there is no such correspondence for BCS-SO* for Case 1 (Figure 14(a)(vi)) because of the larger $\epsilon$ values used in the termination criterion to avoid excessive iterations. The correspondence of larger error bars with suboptimal reconstruction for BCS-SO is not so obvious, which may lead to false confidence in the reconstruction. Therefore, the effective uncertainty quantification of the BCS-SO* method is an obvious advantage. Because the sample variance of the measurements $\text{var}[\mathbf{y}]$ and $\text{var}[\mathbf{y}_d]$, and the corresponding chosen constant prediction error variance $\sigma^2 = \text{var}[\mathbf{y}] \times 0.1$ and $\sigma^2 = \text{var}[\mathbf{y}_d] \times 0.1$ for BCS-B-F, are relatively large in this example, the average error bars for BCS-B-F are larger than those of BCS-SO*; however, it still cannot quantify the signal reconstruction confidence as effectively as BCS-SO*.

For values of *K* that are large enough to give good reconstructions, the BCS-SO* method provides the best overall performance among all methods considering reconstruction error, sparsity and uncertainty quantification of the results.

## 5. CONCLUDING REMARKS

In this paper, the improvement and application of the BCS reconstruction method for compressed signals is studied. This reconstruction method uses a sparse Bayesian learning framework to estimate the sparse signal coefficients in some orthonormal basis. We show that when the number of measurements is much smaller than the length of the discrete-time signal, BCS reconstruction lacks robustness.

Based on these studies, an improved method is developed which uses stochastic optimization and successive relaxation for optimization of the hyper-parameters in sparse Bayesian learning in order to reduce the chance of suboptimal signal reconstructions. Both synthetic signals and real acceleration data from a bridge SHM system are employed to validate the developed method. It is demonstrated that overall the proposed BCS algorithm has a better performance than state-of-the-art BCS algorithms. The signal reconstruction robustness increases significantly and the reconstruction confidence is quantified effectively by the posterior error bars. The noise robustness ability of the modified BCS reconstruction method is also investigated and it is shown to exhibit strong robustness.

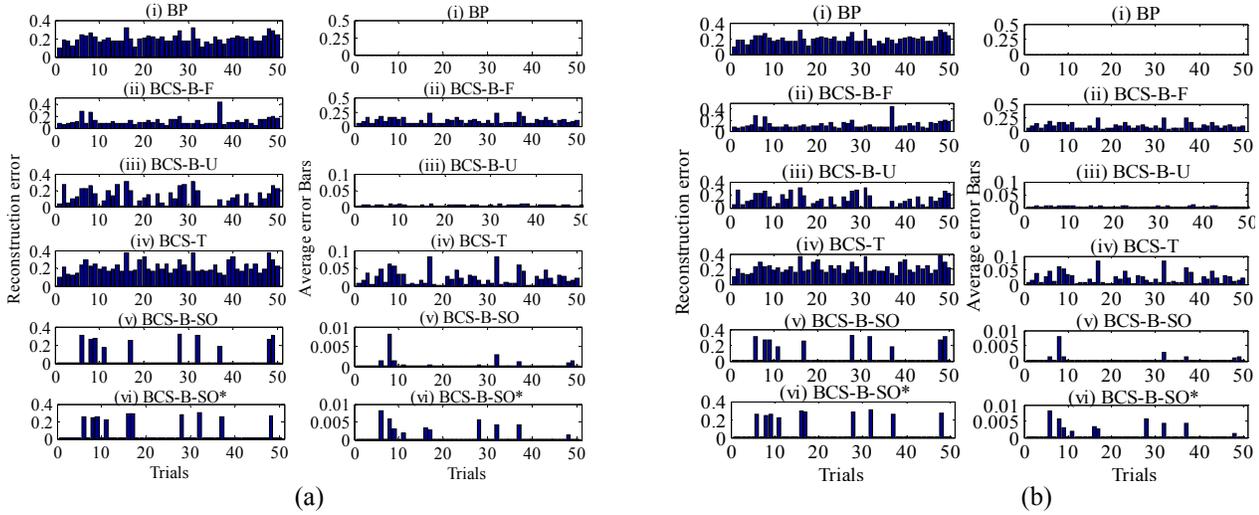

**Figure 14** The reconstruction error and the corresponding average error bars over the first 50 segments of real SHM acceleration data in Figure 9, *K*=200: Results of six algorithms for (a) Case 1; (b) Case 2.

In future studies, it would be useful to explore a Bayesian statistical approach for robust diagnosis of sub-optimal signal reconstructions and to propose algorithms for "healing" them to increase the utility of the BCS technique for SHM signals.

## ACKNOWLEDGMENTS

This work was supported by the U.S. National Science Foundation under award number EAR-0941374 to the California Institute of Technology. This support is gratefully acknowledged by the first two authors. This research is also supported by grants from the National Natural Science Foundation of China (NSFC grant nos. 50538020, 50278029, and 50525823), which supported the fourth author and this support is gratefully acknowledged.

## REFERENCES

Bao, Y.Q., Beck, J.L. and Li, H. (2011), Compressive Sampling for Accelerometer Signals in Structural Health Monitoring, *Structural Health Monitoring*, **10**(3), 235-246.

Beck, J.L. (2010), Bayesian System Identification Based on Probability Logic, *Structural Control and Health Monitoring*, **17**(7), 825-847.

Beck, J.L. and Au, S.K. and Vanik, M.W. (2001), Monitoring Structural Health Using a Probabilistic Measure, *Computer-Aided Civil and Infrastructure Engineering,* 16 (1), 1-11.

Beck, J.L. and Katafygiotis, L.S. (1998), Updating Models and Their Uncertainties I: Bayesian Statistical Framework, *Journal of Engineering Mechanics*, **124**, 455-461.

Candes, E.J. and Wakin, M. (2008), An Introduction to Compressive Sampling, *IEEE Signal Processing Magazine,* **25**(2), 21- 30.

Candes, E.J., Romberg, J. and Tao, T. (2006), Robust Uncertainty Principles: Exact Signal Reconstruction from Highly Incomplete Frequency Information, *IEEE Transactions on Information Theory*, **52**(2), 489-509.

Chen, S.S., Donoho, D.L. and Saunders, M.A. (1999), Atomic Decomposition by Basis Pursuit, *SIAM Journal on Scientific and Statistical Computing*, **20**(1), 33-61.

Donoho, D. (2006), Compressed sensing, *IEEE Transactions on Information Theory*, 52(4), 1289-1306.

Gangone, M.V., Whelan, M.J., and Janoyan, K.D. (2011), Wireless Monitoring of a Multi-Span Bridge Superstructure for Diagnostic Load Testing and System Identification, *Computer-Aided Civil and Infrastructure Engineering*, 26(7), 560-579.

Jaynes E.T. (1957), Information Theory and Statistical Mechanics. *Physical Review*, **106**, 620–630.

Ji, S., Xue, Y. and Carin, L. (2008), Bayesian Compressive Sensing, *IEEE Transactions on Signal Processing*, **56**(6), 2346 - 2356.

Jiang, X. and Adeli, H. (2007), Pseudospectra, MUSIC, and Dynamic Wavelet Neural Network for Damage Detection of Highrise Buildings, *International Journal for Numerical Methods in Engineering*, **71**(5), 606-629.

Lynch, J.P. (2007), An Overview of Wireless Structural Health Monitoring for Civil Structures, P*hilosophical Transactions of the Royal Society A*, **365**: 345-372.

Lynch, J.P., Sundararajan, A., Law, K.H., Kiremidjian, A. S. and Carryer, E. (2003), Power-efficient Data Management for a Wireless Structural Monitoring System, in *Proceedings of the 4th International Workshop on Structural Health Monitoring, Stanford*, CA, 1177-1184.

Sohn, H., Farrar, C.R., Hemez, F.M., Shunk, D.D., Stinemates, D.W., and Nadler, B.R. (2003), A Review of Structural Health Monitoring Literature: 1996-2001, *Los Alamos National Laboratory* Rep. No. LA-13976-MS, Los Alamos National Laboratory, Los Alamos, N.M.

Tipping, M.E. (2001), Sparse Bayesian learning and the relevance vector machine, *Journal of Machine Learning Research*, **1**, 211-244.

Tipping, M.E. and Faul, A.C. (2003), Fast Marginal Likelihood Maximisation for Sparse Bayesian Models, *Proceedings of the Ninth International Workshop on Artificial Intelligence and Statistics*, Key West, FL.

Tropp, J. A. and Gilbert, A. C. (2007), Signal Recovery from Random Measurements via Orthogonal Matching Pursuit, *IEEE Transactions on Information Theory*, **53**(12), 4655-4666.

Vanik, M.W. and Beck, J.L. and Au, S.K. (2000), Bayesian Probabilistic Approach to Structural Health Monitoring, *Journal of Engineering Mechanics,* 126 (7), 738-745.

Wipf, D. and Rao, B. (2006), Comparing the Effects of Different Weight Distributions on Finding Sparse Representations, in *Advances in Neural Information Processing Systems*, **18**,1521-1528.

Xu, N., Rangwala, S., Chintalapudi, K., Ganesan, D., Broad, A., Govindan, R. and Estrin, D. (2004), A Wireless Sensor Network for Structural Monitoring, in *Proceedings of the ACM Conference on Embedded Networked Sensor Systems*. Baltimore, MD, USA.

Yoo, J., Turnes, C., Nakamura, E., Le, C., Becker, S., Sovero, E., Wakin, M., Grant, M., Romberg, J., Emami-Neyestanak, A. and Candès, E. (2012), A Compressed Sensing Parameter Extraction Platform for Radar Pulse Signal Acquisition, *IEEE Journal on Emerging and Selected Topics in Circuits and Systems*, **2**(3): 626-638.

## APPENDIX A. PREDICTION ERROR VARIANCE OPTIMIZATION

For the prediction error variance $\sigma^2$ and parameters $a$ and $b$ in Eqs. (14-16), re-estimation equations can be derived as follows.

From the Woodbury inversion identity,

$$(\sigma^2 \mathbf{I} + \mathbf{\Theta}\mathbf{A}^{-1}\mathbf{\Theta}^T)^{-1} = \sigma^{-2}\mathbf{I} - \sigma^{-2}\mathbf{\Theta}\mathbf{\Sigma}\mathbf{\Theta}^T\sigma^{-2} \quad (29)$$

Then:

$$\begin{aligned}\mathbf{y}^T\mathbf{C}^{-1}\mathbf{y} &= \mathbf{y}^T(\sigma^2\mathbf{I} + \mathbf{\Theta}\mathbf{A}^{-1}\mathbf{\Theta}^T)^{-1}\mathbf{y} \\ &= \sigma^{-2}\mathbf{y}^T\mathbf{y} - \sigma^{-4}\mathbf{y}^T\mathbf{\Theta}\mathbf{\Sigma}\mathbf{\Theta}^T\mathbf{y} \\ &= \sigma^{-2}\mathbf{y}^T(\mathbf{y} - \mathbf{\Theta}\boldsymbol{\mu}) \\ &= \sigma^{-2}(\mathbf{y} - \mathbf{\Theta}\boldsymbol{\mu})^T(\mathbf{y} - \mathbf{\Theta}\boldsymbol{\mu}) + \sigma^{-2}\mathbf{y}^T\mathbf{\Theta}\boldsymbol{\mu} - \sigma^{-2}\boldsymbol{\mu}^T\mathbf{\Theta}^T\mathbf{\Theta}\boldsymbol{\mu} \\ &= \sigma^{-2}\|\mathbf{y} - \mathbf{\Theta}\boldsymbol{\mu}\|_2^2 + \boldsymbol{\mu}^T\mathbf{\Sigma}^{-1}\boldsymbol{\mu} - \sigma^{-2}\boldsymbol{\mu}^T\mathbf{\Theta}^T\mathbf{\Theta}\boldsymbol{\mu} \\ &= \sigma^{-2}\|\mathbf{y} - \mathbf{\Theta}\boldsymbol{\mu}\|_2^2 + \boldsymbol{\mu}^T\mathbf{A}\boldsymbol{\mu}\end{aligned} \quad (30)$$

Therefore, the log of (12) is expressed as (using a determinant identity for $|\mathbf{C}|$ in the second equality (Tipping, 2001; Tipping and Faul, 2003)):

$$\begin{aligned}\mathcal{L} &= -\frac{1}{2}K\log 2\pi - \frac{1}{2}\log|\mathbf{C}| - \frac{1}{2}\mathbf{y}^T\mathbf{C}^{-1}\mathbf{y} + a\log b \\ &\quad -\log\Gamma(a) + (a-1)\log\sigma^{-2} - b\sigma^{-2} \\ &= -\frac{1}{2}K\log 2\pi + \frac{1}{2}(\log|\mathbf{A}| - K\log\sigma^2 + \log|\mathbf{\Sigma}|) \\ &\quad -\frac{1}{2}(\sigma^{-2}\|\mathbf{y} - \mathbf{\Theta}\boldsymbol{\mu}\|_2^2 + \boldsymbol{\mu}^T\mathbf{A}\boldsymbol{\mu}) + a\log b - \log\Gamma(a) \\ &\quad +(a-1)\log\sigma^{-2} - b\sigma^{-2}\end{aligned} \quad (31)$$

The derivative of $\mathcal{L}$ with respect to $\sigma^{-2}$ is then:

$$\begin{aligned}\frac{\partial \mathcal{L}}{\partial \sigma^{-2}} &= \frac{1}{2}K\sigma^2 + \frac{1}{2}\frac{\partial \log|\mathbf{\Sigma}|}{\partial \sigma^{-2}} - \frac{1}{2}\|\mathbf{y} - \mathbf{\Theta}\boldsymbol{\mu}\|_2^2 + (a-1)\sigma^2 - b \\ &= \frac{1}{2}K\sigma^2 - \frac{1}{2}\sigma^2\sum_{n=1}^N(1 - \alpha_n\mathbf{\Sigma}_{nn}) - \frac{1}{2}\|\mathbf{y} - \mathbf{\Theta}\boldsymbol{\mu}\|_2^2 + \\ &\quad (a-1)\sigma^2 - b\end{aligned} \quad (32)$$

using

$$\begin{aligned}\frac{\partial \log|\mathbf{\Sigma}|}{\partial \sigma^{-2}} &= -\text{tr}(\mathbf{\Sigma}\mathbf{\Theta}^T\mathbf{\Theta}) \\ &= -\text{tr}(\sigma^2\mathbf{\Sigma}\mathbf{\Sigma}^{-1} - \sigma^2\mathbf{\Sigma}\mathbf{A}) \\ &= -\sigma^2\sum_{n=1}^N(1 - \alpha_n\mathbf{\Sigma}_{nn})\end{aligned} \quad (33)$$

Setting equation (32) to zero gives an equation for updating:

$$(\sigma^2)^{[j+1]} = \frac{\left\|\mathbf{y} - \mathbf{\Theta}\boldsymbol{\mu}^{[j]}\right\|^2 + 2b^{[j]}}{K - \sum_{n=1}^{N_j}\left(1 - \alpha_n^{[j]}\mathbf{\Sigma}_{nn}^{[j]}\right) + 2(a^{[j]} - 1)} \quad (34)$$

For updating the parameter $b$,

$$\frac{d\mathcal{L}}{db} = \frac{a}{b} - \sigma^{-2} = 0 \quad (35)$$

Then re-estimate $b$ by:

$$b^{[j+1]} = a^{[j]}(\sigma^2)^{[j]} \quad (36)$$

Finally, maximizing (31) with respect to $a$, gives:

$$\frac{d\mathcal{L}}{da} = \log b - \psi(a) - \log \sigma^2 = 0 \quad (37)$$

where $\psi(a)$ is the Digamma function. The numerical equation for re-estimating $a$ for the $(j+1)^{th}$ iteration is given by:

$$\log b^{[j+1]} - \psi(a^{[j+1]}) - \log(\sigma^2)^{[j+1]} = 0 \quad (38)$$

It is seen that the optimization of $a$ cannot converge to a finite value, so we fix $a = 1$.

Note that if we substitute $b = a\sigma^2$ into (31), the optimal estimate of $\hat{\sigma}^2$ is:

$$\hat{\sigma}^2 = \frac{\|\mathbf{y} - \mathbf{\Theta}\mathbf{\mu}\|_2^2}{K - \sum_{n=1}^{N}(1 - \alpha_n \Sigma_{nn}) - 2} \quad (39)$$

This is consistent with an iterative solution for $\hat{\sigma}^2$ by iterating between (32) and (34) until convergence. We have found it is better to use (32) and (34) successively in each iteration over all hyperparameters.